\documentstyle[aps]{revtex}  
\begin{document} 
\title{ 
Renormalization Group Improving the Effective Action
} 
\author{ 
David Hochberg$^{+}$, Carmen Molina--Par\'{\i}s$^{++}$,  
Juan P\'erez--Mercader$^{+}$, and Matt Visser$^{+++}$ 
} 
\address{ 
$^{+}$Laboratorio de Astrof\'{\i}sica Espacial y F\'{\i}sica 
Fundamental, Apartado 50727, 28080 Madrid, Spain\\ 
$^{++}$Theoretical Division, Los Alamos National Laboratory, 
Los Alamos, New Mexico 87545, USA\\ 
$^{+++}$Physics Department, Washington University, 
Saint Louis, Missouri 63130-4899, USA\\ 
} 
\date{29 September 1998; Printed \today} 
\maketitle 

{\small {\bf Abstract:} The existence of fluctuations together with
interactions leads to scale-dependence in the couplings of quantum
field theories for the case of quantum fluctuations, and in the
couplings of stochastic systems when the fluctuations are of  thermal
or statistical nature. In both cases the effects of these fluctuations
can be accounted for by solutions of the corresponding renormalization
group equations. We show how the renormalization group equations are
intimately connected with the effective action: given the effective
action we can extract the renormalization group equations;
given the renormalization group equations the effects of these
fluctuations can be included in the classical action by using what is
known as improved perturbation theory (wherein the bare parameters
appearing in tree-level expressions are replaced by their
scale-dependent running forms). The improved action can then be used
to reconstruct the effective action, up to finite renormalizations,
and gradient terms.}
\newcommand{\tr}{\mathop{\mathrm{tr}}} 
\newcommand{\define}{\mathop{\stackrel{\rm def}{=}}}
\def\d{{\mathrm{d}}}
\section{Introduction:} 
The important role of the effective action, and its
specialization to constant fields, the effective potential, as
fundamental constructs in quantum field theory has now been
appreciated for quite some
time~\cite{Perturbative,Non-perturbative,Fujimoto,Gato}.  The main
purpose of the present paper is to show how radiative corrections and
(quantum) fluctuations may be taken into account in the effective
action and effective potential by means of renormalization group
improved perturbation theory, a concept originally introduced long ago
within the context of QED in the landmark work by Gell-Mann and
Low~\cite{Gell-Mann-Low}. (For discussions see
\cite{Bjorken-Drell-1,Bjorken-Drell-2,Landau-Lifshitz,Weinberg-1,Weinberg-2}.)

An expression is said to be {\em renormalization group improved} if
the bare parameters in the corresponding (typically tree-level)
expression are replaced by their scale-dependent running forms
(calculated to some given order in perturbation theory).  Thus, an
improved quantity (whether it be the effective action, effective
potential, a Green function, {\em etc.}) is one which combines
perturbation theory plus the renormalization group. Such improved
quantities are extremely useful since they allow us to go beyond the
strict limitations of ordinary perturbation theory. In practice, an
improved quantity is calculated to a certain number in {\em loops} but
to {\em all} orders in the couplings and includes much more physics
than the same quantity calculated to a given finite order in the
coupling.  In the loop-expansion, we are in effect, summing up
infinite subclasses of diagrams.  Although the concept of improved
perturbation theory has been around since the work of Gell-Mann and
Low~\cite{Gell-Mann-Low}, and has achieved the status of ``folklore'',
it is rather surprising that explicit treatments in the literature are
somewhat sparse~\cite{Weinberg-2}. The purpose of this paper is to
amend this omission and at the same time to show how RG improvement
may be carried out in a straightforward manner by computing the
effective action and the renormalization group equations directly from
the path integral {\em without the need to handle Feynman diagrams}.

We shall show that solving the renormalization group equations and
constructing the improved action is enough to imply a ``reconstruction
theorem'' whereby the most important pieces of the effective action
(the leading logarithms) can be deduced even if the effective action
is itself unknown. (The importance of the leading logarithms in
perturbation theory has been appreciated since the work of the Russian
School in the early days of quantum field theory.
See~\cite{Landau-Lifshitz} for an accessible discussion.) We shall
demonstrate how the renormalization group equations may be obtained by
inspection of the effective action and how renormalization group {\em
improving} the bare action yields the leading logarithms in the
renormalized effective action. (More precisely, the $n$-loop effective
{\em potential} is sufficient to determine almost all the
renormalization group equations to $n$-loops [all but that for the
wavefunction renormalization]. And conversely the renormalization
group equations are enough to enable us to first improve the classical
potential and then to reconstruct the effective potential up to finite
renormalizations.) This reconstruction theorem is very important:
Often it is much easier to calculate the divergent parts of the bare
effective action than it is to calculate the renormalized effective
action. The logarithmically divergent pieces are however enough to
yield the renormalization group equations, which then can be used to
improve the classical action, and finally via the reconstruction
theorem we can deduce the leading logarithms in the effective action
itself. This specifies the effective action up to finite
renormalizations and gradient terms.

Since we want this article to be widely accessible, we shall be as
clear and explicit as possible. We shall first carry out the
demonstration explicitly for $\lambda (\phi^4)_4$ ($\lambda \phi^4$ in
$n=4$ dimensions) at one loop, and then move on to calculate the
improved action for $\lambda (\phi^3)_6$ ($\lambda \phi^3$ in $n=6$
Euclidean spacetime) where the nontrivial wavefunction renormalization
must be taken into account. For completeness we add some comments
concerning the $\lambda (\phi^6)_3$ and $\lambda (\phi^n)_2$ theories.

The plan of this paper is as follows. In the next section we briefly
develop and exhibit the formal unrenormalized expression for the
effective action valid for any scalar field theory and show how the
associated renormalization group equations (RGE) may be obtained from
it by inspection. The recipe is carried out explicitly for $\lambda
\phi^4$ in $n=4$ Euclidean spacetime employing a simple momentum
cutoff scheme that is well suited for treating homogeneous background
fields and computing effective potentials (which are the homogeneous
field limits of the effective action). We extract the RGE's by
inspection, solve them and use their solutions to improve the
classical action.  When we expand out the improved classical action
(to one loop) we can use it to reconstruct the leading logarithm terms
in the effective action (to one loop).  We next turn to the Schwinger
proper-time regularization scheme (useful for calculating the
effective action when the background field is not necessarily
homogeneous) and re-derive the RGE's for $\lambda\phi^4$ as a
check. We then use this formalism to calculate the leading logarithms
in the effective action for $\lambda \phi^3$ in $n=6$, where
non-trivial wavefunction renormalization effects are known to show
up. Once again, we deduce the associated RGE's by inspection and may
use their solutions to improve the classical action. The formalism is
then applied to $\lambda(\phi^6)_3$, where all the $\beta$--functions
vanish to one loop. (This is a specific example of a general
phenomenon that occurs in odd-dimensional spacetimes: even if the
theory is not one-loop finite, it is nevertheless one-loop
non-running. The fact that the coupling constants do not run in
odd-dimensional spacetimes is a reflection of the absence of
logarithmic divergences.) Finally, we look at $\lambda(\phi^n)_2$, and
write down the renormalization group equations, effective potential,
and leading logarithm portion of the effective action for arbitrary
polynomial interactions.
 
We conclude with a summary and brief comments on the use of improved
perturbation theory in the context of stochastic field theories.
 
\section{Preliminaries} 
 
We consider a general renormalizable scalar field theory.  At the
quantum level its generating functional $Z[J]$ can be written
as~\cite{Reviews,Textbooks,Les-Houches-75}
\begin{eqnarray} 
Z[J]= {\cal N} 
\int [{\cal D} \phi] \; 
\exp \left\{ -{S[\phi]\over\hbar}  
             +\int \d^n x {J(x) \phi(x)\over\hbar} \right\}. 
\end{eqnarray} 
where ${\cal N}$ is a normalization factor, and $\hbar$ is (for the 
purposes of this article) a {\em dimensionless} loop-counting 
parameter.  (If we wish to convert the final answers back to physical 
units, the $\hbar$ will of course convert back into the physical 
Planck constant.) By following the standard procedures~\cite{Ramond}, 
we can define 
\begin{equation}
W[J] = +\hbar \ln Z[J],
\end{equation}
the generating functional for the connected correlation functions, and
its Legendre transform $\Gamma[\bar \phi]$, with $J$ and $\bar \phi$
being conjugate variables to each other (in the Legendre transform
sense of the word).  The loop expansion for $\Gamma[\bar \phi]$ is
based on the following background field decomposition of the field
$\phi$
\begin{eqnarray} 
\phi = \bar \phi + \varphi, 
\end{eqnarray} 
in which $\bar\phi[J]$ is considered a mean field which satisfies the 
classical equation of motion with source $J$, 
\begin{eqnarray}  
\left.\left(  
{ \delta S \over \delta \phi }   
\right)\right|_{\bar \phi[J]}= J, 
\end{eqnarray}  
and $\varphi$ is the quantum fluctuation of the field about the given
background. It is usually assumed that this equation has unique
solutions $\bar\phi[J]$, at least for small $J$, and further that for
vanishing source $J=0$, the unique solution is the constant zero mean
field $\bar \phi =0$. This certainly is valid for a symmetric vacuum
in scalar field theories, but it is not an essential aspect of the
formalism.  For instance in gravity, zero source corresponds to flat
Minkowski space, $g_{\mu\nu} =\eta_{\mu\nu}\neq0$, while in
sigma-models zero source corresponds to constant position on the field
manifold.

As is well known, the calculation of the first quantum correction to
the effective action reduces to evaluating a Gaussian functional
integral. The normalization factor turns out to be important in that
it will cancel out the divergent vacuum energy density ({\em i.e.},
the cosmological constant term) from the effective action.  From the
normalization condition
\begin{eqnarray} 
Z[0]&=&1 
\qquad \rightarrow \qquad 
{\cal N}^{-1} =  
\int [{\cal D} \phi] \; \exp \left\{ -{S[\phi]\over\hbar} \right\}, 
\end{eqnarray} 
we derive the following equation for the generating functional of the 
connected Green functions $W[J]$ 
\begin{equation} 
W[J] = W_0[J]+\hbar W_1[J] + O(\hbar^2), 
\end{equation} 
with 
\begin{equation}  
W_0[J] =  
-\left\{ S[\bar\phi[J]] - S[\bar\phi[0]]\right\} 
+  \int \d^n x \; J(x) \bar\phi(x),  
\end{equation} 
and 
\begin{equation} 
W_1[J]= \sqrt{\det S_2(\bar \phi[0]) \over \det S_2(\bar \phi[J]) } = 
- {1\over 2} \left\{ \tr \; \ln {S_2(\bar \phi[J]) \over \mu^2_\infty}
- \tr \; \ln {S_2(\bar \phi[0]) \over \mu^2_\infty}
\right\}. 
\end{equation} 
Here $\bar\phi[J]$ is the solution to the classical equations of
motion with source $J$, and $\bar\phi[0]$ the classical solution with
no source, that is
\begin{eqnarray}
\label{E:phi-bar-J}  
\left.\left( 
{ {\delta S} \over {\delta\phi}} 
\right)\right|_{\bar\phi [J]} = J, 
\qquad {\rm and} \qquad
\left.\left( 
{ {\delta S} \over {\delta\phi}} 
\right)\right|_{\bar\phi [0]} = 0. 
\end{eqnarray}  
Furthermore 
\begin{eqnarray} 
S_2(x_1,x_2;\bar \phi)= \left( {{\delta^2 S[\phi=\bar \phi]}\over  
{\delta \phi(x_1) \delta \phi(x_2)}} \right),
\end{eqnarray} 
and we have introduced an arbitrary dimensional scale factor
$\mu_\infty$ to keep the argument of the logarithm dimensionless. This
$\mu_\infty$ does not necessarily have anything to do with the running
scale that will be introduced later; it has dimensions of mass.

In a symmetric vacuum we typically have $S[\bar \phi[0]] = 0$, unless 
there is an explicit cosmological constant.  To this order we also 
have for the effective action 
\begin{eqnarray} 
\Gamma[\bar \phi]
&\equiv&-W[J]+\int \d^nx \; J(x) \bar\phi(x) 
= -W_0[J]+\int \d^nx \; J(x) \bar\phi(x)- \hbar W_1[J] +O(\hbar^2), 
\end{eqnarray} 
so that we can write 
\begin{equation} 
\label{E:effaction}
\Gamma[\bar \phi]=\Gamma_0[\bar \phi]+\hbar \Gamma_1[\bar \phi] 
+O(\hbar^2), 
\end{equation} 
with 
\begin{equation} 
\Gamma_0[\bar \phi]=S[\bar \phi]-S[\bar\phi[0]], 
\end{equation} 
and
\begin{equation} 
\Gamma_1[\bar \phi]=-W_1[J[\bar \phi]], 
\end{equation} 
so that
\begin{equation} 
\Gamma_1[\bar \phi]= 
{1\over 2} \left\{ \tr \; \ln {S_2(\bar \phi) \over \mu^2_\infty}
- \tr \; \ln {S_2(\bar \phi[0]) \over \mu^2_\infty}
\right\}. 
\end{equation} 
Even if an explicit cosmological constant is present in the tree level
action, it automatically drops out of the one-loop effective action.
The ``$\tr$'' in the expression for $\Gamma_1[\bar \phi]$ involves an
integration over both $x_1$ and $x_2$ as well as an identification of
these two points by means of a delta function $\delta^n (x_1,x_2)$.

If we were working to higher than one-loop order we would have to be
careful to define $\bar \phi[J]$ via
\begin{equation}
\bar\phi[J,x] = {\delta W[J] \over \delta J(x)},
\end{equation}
which implies
\begin{eqnarray}
\label{E:phi-bar-J2}  
\left.\left( 
{ {\delta \Gamma} \over {\delta\phi}} 
\right)\right|_{\bar\phi [J]} = J.
\end{eqnarray}  
Since the explicit calculations in this article are all to one-loop
order, equation (\ref{E:phi-bar-J}) is sufficient for our purposes.

We know that as we shall be carrying out improved perturbation theory
later, this Gaussian (or one-loop) approximation is sufficient.  By
this we mean that the renormalization group equations will give us the
(one-loop) improved and renormalized theory to all orders in the
coupling $\lambda$ {\em in the leading log approximation}.  (In more
general situations there could be many coupling constants.)  The first
important thing to notice is the fact that the quantity $\Gamma_1[\bar
\phi]$ is divergent, and we will need to analyze its divergences. This
will be done in the next section. The second important remark is the
fact that the bare theory does not depend on the arbitrary scale
introduced by the renormalization scheme; therefore we shall be able
to derive the renormalization group equations from the identity
\begin{eqnarray}\label{E:rge} 
\mu \; {{\d \; \Gamma[\bar \phi]} \over {\d\mu}} = 0 = 
\mu \; {{\d \; S[\bar \phi]} \over {\d\mu}} + \hbar \mu  
{{\d \; \Gamma_1[\bar \phi]} \over {\d\mu}} 
+O(\hbar^2). 
\end{eqnarray} 
This equation yields a polynomial in the background field whose 
coefficients are precisely the RGE's.  Although we shall be using 
quantum scalar field theory to illustrate how the mechanism of 
improved perturbation theory works, it is important to keep in mind 
that this latter identity (\ref{E:rge}) holds for {\em all} 
renormalizable field theories, of both the quantum and stochastic 
varieties, and is independent of the {\em nature} of the fluctuations 
that drive the scale-dependence of the couplings appearing in the 
theory.  This is because the effective action (or effective potential) 
is always a renormalization group invariant.  Of course, when applied 
to a general stochastic field theory, the loop counting parameter will 
no longer be $\hbar$, but will be instead the amplitude of the noise 
two-point function.  We shall come back to this important point in our 
final section. 
 
\section{Homogeneous background fields: Regularization} 
 
This section closely follows the two basic early references that treat 
the study of the effective potential, namely\cite{Fujimoto} and 
\cite{Gato}, in which the extraction of the RGE's directly from the 
effective action (or potential) was pointed out and used for the very 
first time.  The limit of the effective action for constant fields 
$\bar \phi(x) = \phi_0$ is the effective potential and generates the 
connected one-particle irreducible Feynman graphs for zero external 
momentum. Since $\lambda \phi^4$ in $n=4$ dimensions has no 
wavefunction renormalization at one loop, we shall consider the 
homogeneous field limit and calculate the effective potential in a 
simple way using a momentum cutoff procedure.  Notice the fact that as 
the background field is homogeneous, the only divergences showing up 
in the one-loop-contribution to the effective action come from the 
effective potential term, and not the kinetic piece.

The specific scalar Lagrangian we are considering is a $\lambda 
\phi^4$ theory in Euclidean $n$-dimensional spacetime is s
\begin{eqnarray} 
{\cal L} = {1 \over 2} \partial_\mu \phi \, \partial^\mu \phi 
+ {1 \over 2} m^2 \phi^2 
+ {\lambda \over {4!}} \phi^4, 
\end{eqnarray} 
and the corresponding classical action is given by 
\begin{eqnarray}\label{E:action4} 
S[\phi]= \int \d^n x \;  \left[
{1 \over 2} \partial_\mu \phi(x) \, \partial^\mu \phi(x) + {1 \over 2} m^2 
\phi^2(x) + {\lambda \over {4!}} \phi^4(x) \right]. 
\end{eqnarray}

We first calculate $S_2(x_1,x_2;\bar \phi)$ for the action 
(\ref{E:action4}) to obtain 
\begin{equation} 
S_2(x_1,x_2;\bar \phi)= 
\left[- \partial_\mu \partial^\mu + m^2 + 
{\lambda \over 2} {\bar \phi}^2\right] \; \delta^n (x_1,x_2), 
\end{equation} 
and 
\begin{equation} 
S_2(x_1,x_2;\bar \phi=0)= 
[- \partial_\mu \partial^\mu + m^2 ] \; \delta^n (x_1,x_2) . 
\end{equation} 
The one-loop contribution to the effective action is therefore 
\begin{eqnarray} 
\Gamma_1[\bar \phi]&=& {1 \over 2} \;  \tr \;    
\ln \left\{ 
{[- \partial_\mu\partial^\mu + m^2 + 
 {\lambda \over 2} {\bar \phi}^2 ] \delta^n (x_1,x_2) 
\over
\mu^2_\infty}
\right\} 
- 
 {1 \over 2} \;  \tr \;    
\ln \left\{ 
{[- \partial_\mu\partial^\mu + m^2] \delta^n (x_1,x_2) 
\over
\mu^2_\infty}
\right\}. 
\end{eqnarray} 
 
A simple integral representation for the effective action may be had 
by going to a momentum representation for the operators and using the 
homogeneity of the background field: 
\begin{eqnarray} 
\Gamma_1[\bar \phi]&=& 
{1 \over 2} \left\{ \tr \; \ln {S_2(\bar \phi)  \over \mu^2_\infty}
- \tr \; \ln {S_2(\bar \phi =0)  \over \mu^2_\infty}
\right\} 
\\ \nonumber  
&=& 
{1 \over 2} \int \d^n x \; \langle x |  
\ln {S_2(\bar \phi)  \over \mu^2_\infty}
-  \ln {S_2(\bar \phi = 0)  \over \mu^2_\infty} |x  \rangle 
\\ \nonumber  
&=& 
{1 \over 2} \int \d^n x \;  \int \d^n q_1 \;  \int \d^n q_2 \;  
\langle x | q_1 \rangle  \; 
\left\langle q_1 \left| 
\ln {S_2(\bar \phi)  \over \mu^2_\infty} - 
\ln {S_2(\bar \phi = 0)  \over \mu^2_\infty} 
\right| q_2 \right\rangle \;
\langle q_2 | x \rangle.
\end{eqnarray} 
By making use of the definition of $S_2[\bar \phi]$ in momentum 
representation, we can write  
\begin{eqnarray}
\label{E:homogeneous}
\Gamma_1[\bar \phi]&=& 
{1 \over 2} \int \d^n x \;  \int \d^n q_1 \;  \int \d^n q_2 \;  
\langle x | q_1 \rangle   
\left\{ 
\ln  \left[{ q_1^2 + m^2 + 
{\lambda \over 2} {\bar \phi}^2 \over \mu^2_\infty}\right]  
- \ln  \left[{q_1^2 + m^2 \over \mu^2_\infty}\right]  
\right\} \delta^n (q_1,q_2) 
\langle q_2 | x \rangle   
\nonumber\\  
&=& 
{1 \over 2} \int \d^n x \;  \int \d^n q \;
\langle x | q \rangle  
\left\{ 
\ln  \left[{ q^2 + m^2 + 
{\lambda \over 2} {\bar \phi}^2  \over \mu^2_\infty} \right]  
- \ln  \left[{q^2 + m^2 \over \mu^2_\infty}\right]  
\right\} 
\langle q | x \rangle   
\nonumber\\ 
&=& 
{1 \over 2} \int \d^n x \;  \int {{\d^n q} \over {{(2 \pi)}^n}} 
 \left\{ \ln  \left[{q^2 + m^2 + 
{\lambda \over 2} {\bar \phi}^2 \over \mu^2_\infty}\right]  
- \ln \left[{q^2 + m^2\over \mu^2_\infty}\right]
\right\}. 
\end{eqnarray} 
We have to be careful here.  The operator $S_2(\bar \phi)$ is only
diagonal in momentum representation {\em if $\bar \phi$ is a
homogeneous field}. Otherwise $\bar \Phi(x) |q_2\rangle$ will not be
an eigenvector of the momentum operator. (Here $\bar \Phi$ is the
operator corresponding to the classical mean field $\bar \phi$.) Only
in the case $\bar \phi(x) =\bar \phi$ do we have the identity $\bar
\Phi(x) |q_2\rangle=\bar \phi|q_2\rangle$. Nevertheless, there is a
lucky accident for fewer than six dimensions: blindly replacing
$\bar\phi$ by $\bar\phi(x)$ in the above (\ref{E:homogeneous})
produces only a {\em finite} error (with terms proportional to
gradients of $\phi$, actually $O(\phi^2(\partial\phi)^2)$) and
accidentally gives the correct divergent terms for the effective
action. For renormalizable theories in fewer than six dimensions this
procedure (accidentally) leads to the correct renormalization group
equations, and is in error only insofar as it drops all (finite)
gradient terms from the effective action. We will have more to say about
this point later on in the discussion, when we analyze the
wavefunction renormalization in $\lambda(\phi^3)_6$.

Let us now we consider the case $n=4$, and define the integral 
\begin{eqnarray}\label{E:integral} 
{\cal I}(u^2)\stackrel{\rm def}{=}  
\int {{\d^4 q} \over {{(2 \pi)}^4}} \; 
\ln { (q^2 + u^2) \over \mu^2_\infty}. 
\end{eqnarray} 
 
As (\ref{E:integral}) depends only on the modulus of $q$, it is 
convenient to make the following change of variables, $y=q^2$, and we 
can then write 
\begin{eqnarray} 
\int {\d y} \;  y \; \ln  {(y + u^2) \over \mu^2_\infty} = 
{1 \over 2} u^2 y - 
{1 \over 4} y^2  
- 
{1 \over 2} u^4 \ln{(y + u^2) \over \mu^2_\infty} + 
{1 \over 2} y^2 \ln{(y + u^2) \over \mu^2_\infty}. 
\end{eqnarray}

We are able now to calculate ${\cal I}({\cal M}^2)$ and ${\cal 
I}(m^2)$ explicitly, where we have also introduced a momentum cutoff 
$\Lambda$ to regulate the expressions and render the theory finite 
\begin{eqnarray} 
{\cal I}({\cal M}^2)&=& {1 \over {16 \pi^2}} \left[ 
{1 \over 2} {\cal M}^2 {\Lambda}^2 - 
{1 \over 4} {\Lambda}^4   
- 
{1 \over 2} {\cal M}^4 \ln{({\Lambda}^2 + {\cal M}^2)  \over \mu^2_\infty}
+ 
{1 \over 2} {\Lambda}^4   \ln{({\Lambda}^2 + {\cal M}^2)  \over \mu^2_\infty}
+ {1 \over 2} {\cal M}^4 \ln{ {\cal M}^2 \over \mu^2_\infty} \right] 
\\ \nonumber 
{\cal I}(m^2)&=& {1 \over {16 \pi^2}} \left[ 
{1 \over 2} m^2 {\Lambda}^2 - 
{1 \over 4} {\Lambda}^4   
- 
{1 \over 2} m^4 \ln{({\Lambda}^2 + m^2)  \over \mu^2_\infty}
+ 
{1 \over 2} {\Lambda}^4   \ln{({\Lambda}^2 + m^2)  \over \mu^2_\infty}
+ {1 \over 2} m^4 \ln {m^2 \over \mu^2_\infty}\right],
\end{eqnarray} 
where we have defined  
\begin{eqnarray} 
{\cal M}^2 \stackrel{\rm def}{=}   m^2 + 
{\lambda \over 2} {\bar \phi}^2. 
\end{eqnarray}

We are interested in separating the finite pieces from the divergent 
ones in $\Gamma_1[\bar \phi]$ (the divergent terms will be taken care 
of in the renormalization procedure chosen; see the following 
section), therefore we must consider limits such as
\begin{eqnarray} 
\lim_{\Lambda \rightarrow + \infty} {\cal I}({\cal M}^2). 
\end{eqnarray} 
By carrying out an expansion in powers of $1/\Lambda$ we obtain 
\begin{eqnarray} 
({16 \pi^2}) 
\lim_{\Lambda \rightarrow + \infty} {\cal I}({\cal M}^2) &=& 
{1 \over 2} {\cal M}^2 {\Lambda}^2 - 
{1 \over 4} {\Lambda}^4   
-{1 \over 2} {\cal M}^4 \ln \left[ {\Lambda^2 \over \mu^2_\infty} \left( 
1 + {{\cal M}^2 \over \Lambda^2} 
\right) 
\right] 
+ 
{1 \over 2} {\Lambda}^4   \ln \left[ {\Lambda^2 \over \mu^2_\infty} \left( 
1 + {{\cal M}^2 \over \Lambda^2} 
\right) 
\right] 
+ {1 \over 2} {\cal M}^4 \ln{{\cal M}^2  \over \mu^2_\infty}
\nonumber\\ 
&=& 
{1 \over 2} {\cal M}^2 {\Lambda}^2 - 
{1 \over 4} {\Lambda}^4   
+( \Lambda^4  - {\cal M}^4) \ln{\Lambda  \over \mu_\infty}
+ {1 \over 2}( \Lambda^4  - {\cal M}^4) 
\ln \left( 
1 + {{\cal M}^2 \over \Lambda^2} 
\right) 
+ {1 \over 2} {\cal M}^4 \ln{{\cal M}^2  \over \mu^2_\infty}
\nonumber\\ 
&=& 
{1 \over 2} {\cal M}^2 {\Lambda}^2 - 
{1 \over 4} {\Lambda}^4   
+( \Lambda^4  - {\cal M}^4) \ln{\Lambda  \over \mu_\infty}
+ {1 \over 2}( \Lambda^4  - {\cal M}^4) 
\left( 
{{\cal M}^2 \over \Lambda^2} 
- {{\cal M}^4 \over {2 \Lambda^4}} + \dots 
\right) 
+ {1 \over 2} {\cal M}^4 \ln{{\cal M}^2  \over \mu^2_\infty}
\nonumber\\
&=& 
{1 \over 2} {\cal M}^2 {\Lambda}^2 - 
{1 \over 4} {\Lambda}^4   
+( \Lambda^4  - {\cal M}^4) \ln{\Lambda  \over \mu_\infty}
+{{{\cal M}^2 \Lambda^2} \over 2} 
- 
{{\cal M}^4 \over 4} 
+ {1 \over 2} {\cal M}^4 \ln{{\cal M}^2  \over \mu^2_\infty} 
+O(\Lambda^{-2}),  
\end{eqnarray} 
Therefore 
\begin{eqnarray} 
\lim_{\Lambda \rightarrow + \infty} {\cal I}({\cal M}^2) &=& 
{{\Lambda^4} \over {16 \pi^2}} \;  
\left( \ln{\Lambda \over \mu_\infty} - {1 \over 4}   
\right) 
+ {{{\cal M}^2 \Lambda^2} \over {16 \pi^2}} 
+ 
{{{\cal M}^4} \over {32 \pi^2}} \; \left( \ln 
{{{\cal M}^2} \over {\Lambda^2}}  - {1 \over 2}   
\right) + O(\Lambda^{-2}),
\end{eqnarray} 
and
\begin{eqnarray}
\lim_{\Lambda \rightarrow + \infty} {\cal I}(m^2) &=& 
{{\Lambda^4} \over {16 \pi^2}} \;  
\left( \ln{\Lambda \over \mu_\infty} - {1 \over 4}   
\right) 
+ {{m^2 \Lambda^2} \over {16 \pi^2}} 
+ 
{{m^4} \over {32 \pi^2}} \; \left( \ln 
{{m^2} \over {\Lambda^2}}  - {1 \over 2}   
\right) + O(\Lambda^{-2}). 
\end{eqnarray} 
This means that the regulated one-loop contribution to the  
effective action for $\lambda \phi^4$ in $n=4$ is given by 
\begin{eqnarray}\label{E:regulated} 
\Gamma_1[\bar \phi]&=& {1 \over {32 \pi^2}} \int \d^4 x \;  
\left\{ ({\cal M}^2 -m^2) \Lambda^2 
+ 
{{{\cal M}^4 } \over {2}} \; \left[ \ln 
{{{\cal M}^2 } \over {\Lambda^2}}  - {1 \over 2}   
\right]  
- 
{{m^4} \over {2}} \; \left[ \ln 
{{m^2} \over {\Lambda^2}}  - {1 \over 2}   
\right] + O(\Lambda^{-2}) \right\}. 
\end{eqnarray} 
Note that $\mu_\infty$ has disappeared, as of
course it should, since it was only introduced in the first place to
make the argument of the logarithm dimensionless.

A particularly useful separation between finite and divergent pieces 
is obtained by introducing a new {\em arbitrary} scale $\mu$ and writing 
\begin{equation}
{{\cal M}^2\over\Lambda^2} = {{\cal M}^2\over\mu^2} {\mu^2\over\Lambda^2},
\qquad
\hbox{and}
\qquad
{{m^2}\over\Lambda^2} = {{m^2}\over\mu^2} {\mu^2\over\Lambda^2},
\end{equation}
so that
\begin{eqnarray}\label{E:regulated-2} 
\Gamma_1[\bar \phi]&=& {1 \over {32 \pi^2}} \int \d^4 x \;  
\left\{ ({\cal M}^2 -m^2) \Lambda^2 
+ 
{({{\cal M}^4-m^4) } \over {2}} \; \left[ \ln 
{{{\mu}^2 } \over {\Lambda^2}}  - {1 \over 2}   
\right] + 
{ {\cal M}^4 \over 2} \; \ln 
{ {\cal M}^2  \over {\mu^2}} 
- 
{{m^4} \over {2}} \; \ln 
{{m^2} \over {\mu^2}} + O(\Lambda^{-2}) \right\}. 
\end{eqnarray} 
This $\mu$ is logically independent from the previous $\mu_\infty$,
and is being used for a different purpose; $\mu$ is being used in
order to collect all the divergent contributions in one place to
separate them from the interesting finite pieces of the effective
action.  We have now developed all the basic tools that we need to
proceed to the next section, where we carry out the renormalization.
 
\section{Effective potential and renormalization} 
 
We recall here that the one-loop effective action is given by 
(\ref{E:effaction}) 
\begin{eqnarray} 
\Gamma[\bar \phi]=S[\bar \phi]+ \hbar \;  \Gamma_1[\bar \phi] + O(\hbar^2) 
, 
\end{eqnarray} 
and the one-loop effective {\em potential} is obtained by calculating 
the one-loop effective action for a homogeneous field $\bar \phi = 
\phi_0$, and dividing by the volume of spacetime $\Omega = \int \d^n 
x$. That is 
\begin{eqnarray} 
{\cal V}[\phi_0]  
\stackrel{\rm def}{=}  
{ \Gamma[\bar \phi = \phi_0] \over \Omega}= 
V[\bar \phi = \phi_0] +  
\hbar {\Gamma_1[\bar \phi = \phi_0]\over\Omega} +  
O(\hbar^2) 
, 
\end{eqnarray} 
with  
\begin{eqnarray} 
V[\phi] =   
{m^2 \over 2} \phi^2 + {\lambda \over {4 !}}\phi^4, 
\end{eqnarray} 
the classical potential.  
 
{From} (\ref{E:regulated-2}) we can write 
\begin{eqnarray} 
\label{E:effective-potential:bare} 
{\cal V}[\phi_0] &=&  
V[\phi_0]  
+ {\hbar \over {32 \pi^2}} \left[ 
 \Lambda^2 {\lambda \over 2} \phi_0^2 
+ 
{{({\cal M}_0^4 - m^4)} \over {2}} \; \left( \ln 
{{\mu^2} \over {\Lambda^2}}  - {1 \over 2}   
\right)  
+ 
{{{\cal M}_0^4} \over {2}} \;  \ln 
{{{\cal M}_0^2} \over {\mu^2}}  
- 
{{m^4} \over {2}} \;  \ln 
{{m^2} \over {\mu^2}} 
+O(\Lambda^{-2}) 
\right] + O(\hbar^2)
\\ 
&=& 
{m^2 \over 2} \phi_0^2 + {\lambda \over {4 !}}\phi_0^4 
\nonumber\\
&&
\qquad \; \; 
+ {\hbar \over {32 \pi^2}} \left[ 
 \Lambda^2 {\lambda \over 2} \phi_0^2 
+ 
{{({\cal M}_0^4 - m^4)} \over {2}} \; \left( \ln 
{{\mu^2} \over {\Lambda^2}}  - {1 \over 2}   
\right)  
+ 
{{{\cal M}_0^4} \over {2}} \;  \ln 
{{{\cal M}_0^2} \over {\mu^2}}  
- 
{{m^4} \over {2}} \;  \ln 
{{m^2} \over {\mu^2}}  
+O(\Lambda^{-2}) 
\right] + O(\hbar^2), 
\end{eqnarray} 
with ${\cal M}_0^2=m^2+{\lambda \over 2} {\phi_0}^2$. At this stage
the expressions are all given in terms of bare (unrenormalized)
parameters and the dependence on the cutoff is explicit. We have
introduced the {\em arbitrary} scale $\mu$ to get the cutoff
dependent pieces concentrated in one place. The effective potential is
seen by inspection to be independent of $\mu$.
 
Since we are dealing with a renormalizable theory, we know that we can
absorb the cutoff dependence into the parameters $m$ and $\lambda$.
Because there is no wavefunction renormalization, $Z= 1 + O(\hbar^2)$
to one loop in $\lambda(\phi^4)_4$ field theory, we do not need to
worry about this particular complication. Otherwise we would have
directly gone to investigating the effective action $\Gamma[\phi_0]$.
Specifically, let us write
\begin{eqnarray} 
m^2 &=& m^2(\mu) + \hbar [\delta m^2](\mu) + O(\hbar^2), 
\\ 
\lambda &=& \lambda(\mu) + \hbar [\delta\lambda](\mu) + O(\hbar^2). 
\end{eqnarray} 
For a specific and useful choice of separation into renormalized
parameters and counterterms the classical potential $V[\phi_0]$ may
then be written as
\begin{eqnarray}  
\label{E:counterterms}
V[\phi_0]  &=& 
V[\phi_0,m(\mu),\lambda(\mu)] 
-{\hbar \over {32 \pi^2}} \left[ 
 \Lambda^2 {\lambda \over 2} \phi_0^2 
+ 
{{({\cal M}_0^4 - m^4)} \over {2}} \;  
\left( \ln{{\mu^2} \over {\Lambda^2}} - {1\over2} \right) 
\right] + O(\hbar^2), 
\end{eqnarray} 
where the counter terms have been so chosen so that they will render
the one-loop effective potential finite and cutoff independent.
Notice also the very important fact that in the $O(\hbar)$ term above,
the parameters can with equal facility be taken to be either the bare
ones or the renormalized ones, since the difference will contribute to
the full expression only at $O(\hbar^2)$.
 
We conclude that the effective potential, up to one loop, is given by 
\begin{eqnarray} 
\label{E:effective-potential:partial} 
\label{E:effective-potential:semi} 
{\cal V}[\phi_0] &=&  
V[\phi_0, m(\mu),\lambda(\mu)]  
+ {\hbar \over {64 \pi^2}} \left\{ 
{{\cal M}_0^4} \;  \ln 
{{{\cal M}_0^2} \over {\mu^2}}  
- 
{{m^4}} \;  \ln 
{{m^2} \over {\mu^2}}  
\right\} +O(\hbar^2) 
\\ 
&=&  
{m^2(\mu) \over 2} \phi^2_0 + {\lambda (\mu) \over {4 !}}\phi^4_0  
+ {\hbar \over {64 \pi^2}} \left\{ 
{{{\cal M}_0^4}} \;  \ln {{{\cal M}_0^2} \over {\mu^2}}  
- 
{{m^4}} \;  \ln {{m^2} \over {\mu^2}}  
\right\} + O(\hbar^2). 
\end{eqnarray} 
The parameters $m(\mu)$ and $\lambda (\mu)$ are the renormalized ones, 
whereas the parameter ${\cal M}_0^2 = m^2 + {\lambda \over 2} 
\phi^2_0$ is still written in terms of the bare parameters $m$ and 
$\lambda$.  (This is actually convenient for some purposes!) Of
course, to this order in the loop expansion, it is equally valid (and
considerably more elegant) to write
\begin{equation} 
\label{E:effective-potential:renormalised} 
{\cal V}[\phi_0] =  
{m^2(\mu) \over 2} \phi^2_0 + {\lambda (\mu) \over {4!}}\phi^4_0
+ {\hbar \over {64 \pi^2}} \left\{ 
{{{\cal M}_0^4(\mu)}} \;  \ln {{{\cal M}_0^2(\mu)} \over {\mu^2}}  
- 
{{m^4(\mu)}} \;  \ln {{m^2(\mu)} \over {\mu^2}}  
\right\} + O(\hbar^2). 
\end{equation} 
Perhaps the key ``miracle'' of renormalization is this: although there 
are many occurrences of the parameter $\mu$ on the right hand side of 
this equation, the left hand side is completely independent of 
$\mu$. After all, the above expression is a re-writing of 
equation (\ref{E:effective-potential:bare}). The ``miracle'' is of 
course no miracle: it has been enforced by explicit construction, 
and will be the underpinning of our derivation of the renormalization 
group equations. 

Note that the above represents a particular {\em prescription} for
renormalization. We can always change the division between running
parameters and counterterms in equation (\ref{E:counterterms}) by
arbitrary finite quantities without disturbing the elimination of the
cutoff dependencies. The prescription chosen here has the advantages
of both providing a clean analytic form for the one-loop effective
potential and simultaneously being well adapted to the massless
limit. Indeed if we let the mass go to zero we get
\begin{equation} 
\label{E:effective-potential:massless} 
{\cal V}[\phi_0] =  
{\lambda (\mu) \over {4!}}\phi^4_0
+ {\hbar \over {64 \pi^2}} \left\{ 
{\lambda(\mu)^2 \phi_0^4\over4} 
\;  \ln {\lambda(\mu) \phi_0^2\over {2 \mu^2}}  
\right\} + O(\hbar^2). 
\end{equation} 
Equivalently
\begin{equation} 
\label{E:effective-potential:massless2} 
{\cal V}[\phi_0] =  
{\lambda (\mu) \over {4!}}\phi^4_0 
\left\{
1 + \hbar {3\lambda(\mu) \over {32 \pi^2}} 
\;  \ln {\lambda(\mu) \phi_0^2\over {2 \mu^2}}  
\right\} + O(\hbar^2). 
\end{equation} 
We shall use this massless effective potential when we show how to
reconstruct the one-loop effective potential from the one-loop
renormalization group improved bare potential.

To make the ambiguity under finite renormalizations explicit, suppose
we replace (\ref{E:counterterms}) above by
\begin{eqnarray}  
\label{E:counterterms2}
V[\phi_0]  &=& 
V[\phi_0,m(\mu),\lambda(\mu)] 
-{\hbar \over {32 \pi^2}} \left[ 
 \Lambda^2 {\lambda \over 2} \phi_0^2 
+ 
{{({\cal M}_0^4 - m^4)} \over {2}} \;  
\left( \ln{{\mu^2} \over {\Lambda^2}} - {1\over2} \right) 
\right]
\nonumber\\
&&\qquad
+\hbar \, \epsilon_1 \, m^2(\mu) \, \phi_0^2 
+\hbar \, \epsilon_2 \, \phi_0^4
+ O(\hbar^2), 
\end{eqnarray} 
where $m(\mu)$ and $\lambda(\mu)$ are new and slightly different
renormalized couplings. (They will, however, satisfy the same
renormalization group equations, at least to one loop.) We have also
used dimensional analysis to constrain the ambiguities and place all
the arbitrariness into two dimensionless parameters $\epsilon_1$ and
$\epsilon_2$.  Then the previous analysis continues to hold except
that the effective potential is replaced by
\begin{eqnarray} 
\label{E:effective-potential:renormalised2} 
{\cal V}[\phi_0] &=&  
{m^2(\mu) \over 2} \phi^2_0 + {\lambda (\mu) \over {4!}}\phi^4_0
+ {\hbar \over {64 \pi^2}} \left\{ 
{{{\cal M}_0^4(\mu)}} \;  \ln {{{\cal M}_0^2(\mu)} \over {\mu^2}}  
- 
{{m^4(\mu)}} \;  \ln {{m^2(\mu)} \over {\mu^2}}  
\right\} 
\nonumber\\
&&\qquad
+\hbar \, \epsilon_1 \, m^2(\mu) \, \phi_0^2 
+\hbar \, \epsilon_2 \, \phi_0^4 
+ O(\hbar^2). 
\end{eqnarray} 
This two-parameter ambiguity in the effective potential is an
intrinsic and unavoidable side-effect of renormalization: this
ambiguity may be used to force the effective potential to have certain
simplifying properties, and the choice made previously [in
(\ref{E:counterterms})] was exactly one such choice. Another common
choice, if $m(\mu)\neq0$, is to fix the derivatives of the effective
potential at zero field to be
\begin{eqnarray}
\left.{\d^{(4)} {\cal V} \over \d\phi_0^4}\right|_{\phi_0=0} &=& \lambda(\mu)
\\
\left.{\d^{(2)} {\cal V} \over \d\phi_0^2}\right|_{\phi_0=0} &=& m^2(\mu).
\end{eqnarray}
These renormalization conditions are equivalent to particular choices
of $\epsilon_1$ and $\epsilon_2$. See, for example, pages 453--454 of
Itzykson and Zuber~\cite{Itzykson-Zuber}. If $m(\mu)=0$, a popular
choice is
\begin{eqnarray}
\left.{\d^{(4)} {\cal V} \over \d\phi_0^4}\right|_{\phi_0=\mu} &=& \lambda(\mu).
\end{eqnarray}
See, for example, page 454 of Itzykson and
Zuber~\cite{Itzykson-Zuber}.  We shall eschew such specific choices
and stay with the simple form
(\ref{E:effective-potential:renormalised}) above.

\section{Renormalization by differentiation and subsequent integration} 
 
For comparison, we include here some remarks that we believe to be
helpful, regarding an alternative approach to the divergent structure
of the theory. The main reference for this section is
Weinberg~\cite{Weinberg-2}, (see chapter 16) though related
discussions can be found in many field theory textbooks. In Feynman
diagram language the idea is to differentiate with respect to some
external parameter a sufficient number of times and so render the relevant
integrals finite. These differentiations may be with respect to
external momenta (where they underly the BPHZ renormalization
program), with respect to some particle mass, or (for the case we are
interested in) the derivative may be with respect to the external
field. After differentiation has rendered the integrals finite, the
result is re-integrated an equal number of times.  Each integration
introduces a new constant of integration; these arbitrary constants of
integration {\em are} the counterterms.

 
In our case, ${\cal V}_1$ contains ultraviolet divergences. Since this
theory is perturbatively renormalizable, these divergences can be
absorbed into a renormalization of the parameters of the
theory. ${\cal V}_1$ is divergent, and by power counting, it can be
seen that it is made convergent by differentiating three times with
respect to ${\cal M}^2$. We start by writing
\begin{equation} 
{{\cal V}_1[\phi_0]}={\cal W}_1({\cal M}^2_0)-{\cal W}_1(m^2), 
\end{equation} 
with
\begin{equation}
{\cal W}_1[a] = \int {\d^4q\over(2\pi)^4} \ln {(q^2 + a) \over \mu^2_\infty}.
\end{equation}
Differentiating thrice
\begin{equation} 
{{\d^3 {\cal W}_1[a]}\over {\d a^3}} = 
{1 \over {8 \pi^2}} 
\int \d y {y \over {{(y+a)}^3}} 
= 
{1 \over {32 \pi^2 a}}. 
\end{equation} 
Integrating the previous equation three times, and adding the 
corresponding constants of integration, we obtain 
\begin{eqnarray} 
{\cal W}_1[a]
&=&
{a^2 \over {64 \pi^2}} \ln{a\over\mu^2} 
+ \kappa_1(\mu) a^2 + \kappa_2 a + \kappa_3, 
\end{eqnarray} 
with the $\kappa$'s being divergent coefficients. The dimensional
parameter $\mu$ {\em must} be introduced to keep the argument of the
logarithm dimensionless, and implies a $\mu$ dependence in $\kappa_1$
in such a manner that the left hand side above is independent of
$\mu$. Since the $\mu$ independence of the left hand side holds for
all values of $a$, the same logic implies that $\kappa_2$ and
$\kappa_3$ cannot be functions of $\mu$. (In this formalism we do not
need to keep $\mu$ and $\mu_\infty$ distinct, and we may choose to
conflate them if we wish.) Therefore
\begin{eqnarray} 
{{\cal V}}&=& 
{m^2 \over 2} \phi^2_0 + {\lambda \over {4 !}}\phi^4_0+ 
\hbar\left\{ {\cal W}_1[{\cal M}^2_0]-{\cal W}_1[m^2]\right\} + O(\hbar^2) 
\\ 
&=& 
{m^2 \over 2} \phi^2_0 + {\lambda \over {4 !}}\phi^4_0+ 
{\hbar \over {64 \pi^2}}  
\left\{ 
{\cal M}^4_0 \ln\left({ {\cal M}^2_0\over\mu^2}\right) 
-{m}^4 \ln\left({ {m}^2 \over\mu^2}\right)  
\right\} 
+ \hbar \kappa_1(\mu)  
\left( \lambda m^2 \phi^2_0 + {{\lambda^2 \phi^4_0} \over 4}\right) 
+\hbar\kappa_2 {\lambda \over 2} \phi^2_0  
+ O(\hbar^2) 
\\  
&=& 
{m^2 + 2 \hbar\kappa_1(\mu) m^2 + \hbar \kappa_2\lambda \over 2} \; \phi^2_0  
+ {\lambda + 6 \hbar \kappa_1(\mu) \lambda^2 \over {4 !}} \; \phi^4_0+ 
{\hbar \over {64 \pi^2}}  
\left\{ 
{\cal M}^4_0 \ln\left({ {\cal M}^2_0\over\mu^2}\right) 
-{m}^4 \ln\left({ {m}^2 \over\mu^2}\right) 
\right\} 
+O(\hbar^2). 
\end{eqnarray} 
Everything is here still expressed in terms of unrenormalized 
quantities and cutoff dependent parameters $\kappa_i$.  The theory is 
renormalizable, so that we expect to be able to write $\cal V$ as a 
function of the renormalized parameters, with no explicit divergences, 
though we will just have to live with the presence of the scale $\mu$. 
Define renormalized parameters by 
\begin{eqnarray} 
m^2(\mu)&=&  
m^2 + \hbar[2\lambda m^2 \kappa_1(\mu) + \lambda \kappa_2 ] 
+ O(\hbar^2), 
\\  
\lambda(\mu)&=&  
\lambda + \hbar [6 \lambda^2 \kappa_1(\mu)] + O(\hbar^2). 
\end{eqnarray} 
Remember that the divergences are all buried in the $\kappa_i$'s.
This choice of renormalized parameters is again to some extent
arbitrary, but is a particularly simple one based on keeping the form
of the effective potential invariant. Any other choice of
renormalization prescription will differ from this one by only finite
quantities and will at worst lead to finite renormalization ambiguities
in the effective action. In terms of these particular renormalized
parameters we can write the partially renormalized expression
\begin{eqnarray} 
\label{E:effective-potential:semi2} 
{{\cal V}}&=& 
{m^2(\mu) \over 2} \phi^2_0 + {\lambda(\mu) \over {4 !}}\phi^4_0+ 
{\hbar \over {64 \pi^2}}  
\left\{ 
{\cal M}^4_0 \ln\; {{\cal M}^2_0 \over \mu^2} - m^4 \ln {m^2\over\mu^2}  
\right\} + O(\hbar^2), 
\end{eqnarray} 
where in the $O(\hbar)$ terms we have kept the bare parameters.  Since 
bare and renormalized parameters differ at $O(\hbar)$ this is 
completely equivalent to the expression written completely in terms of 
renormalized parameters 
\begin{eqnarray} 
{{\cal V}}&=& 
{m^2(\mu) \over 2} \phi^2_0 + {\lambda(\mu) \over {4 !}}\phi^4_0+ 
{\hbar \over {64 \pi^2}}  
\left\{ 
{\cal M}^4_0(\mu) \ln\; {{\cal M}^2_0(\mu) \over \mu^2} -  
m^4(\mu) \ln {m^2(\mu)\over\mu^2}  
\right\} + O(\hbar^2). 
\end{eqnarray} 
All the $\mu$ dependencies on the right hand side of the above cancel 
exactly: the left hand side is known by construction to be independent 
of $\mu$.  Finally notice that all these results agree with the 
arguments in the previous section.

\section{Renormalization group equations for the $\lambda \phi^4$-theory} 
 
As we remarked earlier, there is no wavefunction renormalization at 
this order in the loop expansion (for the $\lambda \phi^4$ theory in 
$4$ dimensions), and the condition 
\begin{eqnarray} 
\mu \; {{\d \; \Gamma[\bar \phi]} \over {\d\mu}} = 0  
\qquad \Rightarrow \qquad 
\mu \; {{\d \; {\cal V}[\phi_0]} \over {\d\mu}} = 0. 
\end{eqnarray} 
This can be seen more directly, by noting (at this stage without
proof) that the general form of the renormalized effective action at
one-loop order is
\begin{eqnarray} 
\Gamma[\bar \phi] = 
\int \d^n x \left[ 
{1 \over 2} Z(\mu) \, \partial_\mu \bar \phi \, \partial^\mu \bar \phi 
+ {\cal V} (\bar \phi)
+ O(\hbar \bar\phi^2  \, \partial_\mu \bar \phi \, \partial^\mu \bar \phi )
+ O(\hbar^2)
\right]. 
\end{eqnarray} 
At one-loop order for a $\lambda \phi^4$ theory there is no
wavefunction renormalization. That is $Z(\mu)=1 +O(\hbar^2)$, and
$\bar\phi$ is itself the renormalized and improved field (it does not
depend on the scale $\mu$).
\begin{eqnarray} 
\bar \phi = Z^{1/2}(\mu) \, \phi_0 = \phi_0. 
\end{eqnarray} 
It is now easy to see that the scale independence of the one-loop 
effective action implies (in this particular case) the scale 
independence of the one-loop effective potential, where 
\begin{eqnarray} 
{\cal V}[\phi_0] &=&  
V[\phi_0, \mu]  
+ {\hbar \over {64 \pi^2}}  
\left\{ 
{{{\cal M}_0^4(\mu)}} \;  \ln 
{{{\cal M}_0^2(\mu)} \over {\mu^2}} 
- 
{m^4(\mu)} \ln {m^2(\mu)\over\mu^2}  
\right\} 
+ O(\hbar^2), 
\end{eqnarray} 
and 
\begin{eqnarray} 
V[\phi_0, \mu] = {{m^2 (\mu)} \over 2} \phi_0^2 
+ {{\lambda (\mu)} \over {4!}} \phi_0^4. 
\end{eqnarray} 
Notice that in all the expressions after this one, we mean for $m$ and 
$\lambda$ the renormalized, scale dependent parameters $m(\mu)$, and 
$\lambda (\mu)$. Differentiating, we get 
\begin{eqnarray} 
0&=& \mu {\d \over {\d \mu}} \left( {{m^2} \over 2} \phi_0^2 
+ {\lambda \over {4!}} \phi_0^4 
\right)   
+ 
{\hbar \over {64 \pi^2}}  
\mu {\d \over {\d \mu}} 
\left\{ 
{{\cal M}_0^4(\mu)} \;  \ln 
{{{\cal M}_0^2(\mu)} \over {\mu^2}}  
- 
m^4(\mu) \ln {m^2(\mu)\over\mu^2}  
\right\} 
+O(\hbar^2), 
\end{eqnarray} 
which implies 
\begin{eqnarray} 
\mu {\d \over {\d \mu}} \left( {{m^2} \over 2} \phi_0^2 
+ {\lambda \over {4!}} \phi_0^4 
\right) & =&   
{\hbar \over {32 \pi^2}}  
({\cal M}^4_0  - m^4) + O(\hbar^2) 
\\ 
&=& {\hbar \over {32 \pi^2}} \left(  
{\lambda^2 \over 4} \phi_0^4 + \lambda m^2 \phi_0^2 \right) 
+ O(\hbar^2). 
\end{eqnarray} 
Note that derivatives of ${\cal M}$ and $m$ with respect to $\mu$ on
the right hand side of the two equations above lead to contributions
only at $O(\hbar^2)$ and so can be neglected at the order we are
interested in. Equivalently, we could apply the same differentiation
to the semi-renormalized effective potential, equations
(\ref{E:effective-potential:semi}) or
(\ref{E:effective-potential:semi2}). This derivative technique was
first used in the work of Fujimoto, O'Raifeartaigh, and
Parravicini~\cite{Fujimoto}, and was extensively developed in the work
of Gato, Le\'on, P\'erez--Mercader, and Quir\'os~\cite{Gato}. It is a
very simple and powerful technique that deserves wider use.
 
Finally, we conclude 
\begin{eqnarray}\label{E:match} 
{\phi_0^2 \over 2} \mu {{\d m^2} \over {\d\mu}} 
+ {\phi_0^4 \over {4!}}  
\mu {{\d \lambda} \over {\d\mu}} 
= 
{\hbar \over {128 \pi^2}} {\phi_0^4} \lambda^2 
+{\hbar \over {32 \pi^2}} {\phi_0^2} m^2 \lambda 
+O(\hbar^2). 
\end{eqnarray} 
We can now easily obtain the renormalization group equations for
$m(\mu)$ and $\lambda(\mu)$, and solve them. We need only identify the
terms with the same powers of $\phi_0$. This is because the two sides
of (\ref{E:match}) are self-consistent, as a result of the
renormalizability of the potential. 

(For a non-renormalizable theory [{\em e.g.}, $\lambda(\phi^6)_4$] we
would find different functional forms on the two sides of the
equation, [since ${\cal M}^4$ now contains terms such as $\phi_0^8$]
thus indicating that we had not included enough terms in the classical
potential $V(\phi)$. If we attempt to fix this by adding additional
terms to $V(\phi)$ even higher order terms show up in ${\cal M}^4$ and
we are forced to bootstrap ourselves into a situation where $V(\phi)$
contains {\em all} powers of $\phi$. Viewed as a fundamental theory
the resulting model is generally condemned as being ``non-predictive''
though viewed as an effective theory it is more interesting.)
 
Had we not taken into account the normalization factor for the 
generating functional $Z[J]$, the term involving $m^4$ would not have 
canceled out and we would have had to carry along an additional RGE 
for the vacuum energy density or cosmological constant, while adding a 
tree-level cosmological constant to our bare action.  The careful 
inclusion of the normalization factor $\cal N$ in the partition 
function takes care of subtracting the divergences corresponding to 
the vacuum. 
 
We have then 
\begin{eqnarray}
\label{E:rge-1} 
{\mu} {{\d m^2} \over {\d\mu}}&=& 
{\hbar \over {16 \pi^2}}  m^2 \lambda + O(\hbar^2), 
\\  
\label{E:rge-2}
\mu {{\d \lambda} \over {\d\mu}} &=& 
{{3 \hbar} \over {16 \pi^2}}  {\lambda^2} + O(\hbar^2). 
\end{eqnarray} 
These are the one-loop renormalization group equations for the
theory. It is standard to define~\cite{Nash,Ramond,Collins,Pokorski}
\begin{eqnarray}\label{E:rges} 
\beta \left( \lambda, {m \over \mu} \right) &\stackrel{\rm def}{=}& 
\mu {{\partial \lambda}\over {\partial \mu}}, 
\\  
\gamma_{\mathrm{m}} 
\left( \lambda, {m \over \mu} \right)  
&\stackrel{\rm def}{=}& {1 \over 
2} \mu {{\partial \ln \; m^2}\over {\partial \mu}}, 
\\ 
\gamma_{\mathrm{d}} \left( \lambda, {m \over \mu} \right)  
&\stackrel{\rm def}{=}&  
 {1 \over 2} 
\mu {{\partial \ln \; Z}\over {\partial \mu}}. 
\end{eqnarray} 
Unfortunately, we must warn the reader that various authorities use a
different sign convention for $\gamma_{\mathrm{m}}$.  With these
conventions we obtain
\begin{eqnarray} 
\beta \left( \lambda, {m \over \mu} 
\right) &=& {{3 \hbar \lambda^2}\over {16 \pi^2}} +O(\hbar^2), 
\\ 
\gamma_{\mathrm{m}} \left( \lambda, {m \over \mu} 
\right) &=&  {{\hbar \lambda} \over {32 \pi^2}} +O(\hbar^2), 
\\ 
\gamma_{\mathrm{d}} \left( \lambda, {m \over \mu} 
\right) &=& 0 + O(\hbar^2). 
\end{eqnarray} 
These results agree with Ramond~\cite{Ramond} Chapter $4$, with
Zinn--Justin~\cite{Zinn-Justin} Chapter $11$, and with
Le~Bellac~\cite{Le-Bellac} Chapters $2$, $5$, $6$, and $7$.  (We have
not actually derived $\gamma_{\mathrm{d}}$ at this stage but merely
asserted the result. A proof will be forthcoming shortly.)

The solution of these renormalization group equations 
is straightforward: they are given by 
\begin{eqnarray} 
\lambda &=& \lambda_0 {\left( 
1 - {{3\hbar} \over{16 \pi^2} } \lambda_0 \ln{\mu \over \mu_0} 
 + 0(\hbar^2) \right) }^{-1}, 
\\  
m^2&=& m_0^2  \; 
(\mu/\mu_0)^{{\hbar\lambda_0} \over{16 \pi^2} } \;
\exp(O(\hbar^2)).
\end{eqnarray} 
where $\lambda_0 = \lambda(\mu_0)$, $m_0 = m(\mu_0)$ and 
$\mu_0$ denotes an {\em arbitrary} initial renormalization scale. 

\section{Consistency check on the effective potential}

We now present a simple consistency check to verify that the
effective potential is independent of the renormalization scale
$\mu$. Start with the formula
\begin{equation}
V[\phi_0, \mu] = {{m^2 (\mu)} \over 2} \phi_0^2 
+ {{\lambda (\mu)} \over {4!}} \phi_0^4, 
\end{equation} 
and insert the running couplings we have just derived. Then
\begin{eqnarray} 
V[\phi_0, \mu] 
&=& 
{1 \over 2}  m_0^2  \; 
(\mu/\mu_0)^{{\hbar\lambda_0} \over{16 \pi^2} } \;
\exp(O(\hbar^2)) \; \phi^2_0 
+ \frac{\lambda_0}{4!} {\left( 
1 - {{3\hbar} \over{16 \pi^2} } \lambda_0 \ln{\mu \over \mu_0} 
 + 0(\hbar^2) \right) }^{-1} \phi^4_0. 
\end{eqnarray} 
Expanding this to $O(\hbar^2)$
\begin{eqnarray} 
V[\phi_0, \mu] 
&=& 
{1 \over 2} m_0^2  
{\left( 
1 + {{\hbar} \over{16 \pi^2} } \lambda_0 \ln{\mu \over \mu_0} 
\right)} \phi^2_0 
+ \frac{\lambda_0}{4!} 
{\left( 
1 + {{3\hbar} \over{16 \pi^2} } \lambda_0 \ln{\mu \over \mu_0} 
\right)} \phi^4_0 
+ O(\hbar^2). 
\end{eqnarray} 
This can be re-expressed as
\begin{eqnarray} 
V[\phi_0, \mu] 
&=& 
V[\phi_0, \mu_0] +
 {{\hbar} \over{32 \pi^2} } 
\left\{ 
{1\over2} m_0^2 \lambda_0 \phi_0^2 
+ {1\over8}\lambda_0^2 \phi_0^4
\right\} \ln{\mu^2 \over \mu_0^2} 
+ O(\hbar^2), 
\end{eqnarray} 
or better yet
\begin{eqnarray} 
V[\phi_0, \mu] 
&=& 
V[\phi_0, \mu_0] +
 {{\hbar} \over{64 \pi^2} } 
\left\{ 
{\cal M}_0^4(\mu_0) - m_0^4
\right\} \ln{\mu^2 \over \mu_0^2} 
+ O(\hbar^2). 
\end{eqnarray} 
The point is that after inserting the running couplings, as deduced
from the renormalization group, the $\mu$ dependence in the
renormalized classical potential cancels exactly (to $O(\hbar^2)$) the
$\mu$ dependence in ${\cal V}_1$, so that
\begin{eqnarray} 
{\cal V}[\phi_0] &=&  
V[\phi_0, \mu]  
+ {\hbar \over {64 \pi^2}}  
\left\{ 
{{{\cal M}_0^4(\mu)}} \;  \ln 
{{{\cal M}_0^2(\mu)} \over {\mu^2}} 
- 
{m^4(\mu)} \ln {m^2(\mu)\over\mu^2}  
\right\} 
+ O(\hbar^2), 
\\
&=&  
V[\phi_0, \mu_0]  
+ {\hbar \over {64 \pi^2}}  
\left\{ 
{{{\cal M}_0^4(\mu_0)}} \;  \ln 
{{{\cal M}_0^2(\mu_0)} \over {\mu^2_0}} 
- 
{m^4(\mu_0)} \ln {m^2(\mu_0)\over\mu^2_0}  
\right\} 
+ O(\hbar^2). 
\end{eqnarray} 
This may be a little tedious, but it has the virtue of being explicit, and
verifying the consistency of the whole approach.
 
\section{The improved potential: Reconstruction} 
 
We have just seen how knowledge of the one-loop effective potential
gives the one-loop RGE's; and have verified that the running of the
couplings implied by the RGE's is consistent with the renormalization
scale independence of the effective potential. We shall now point out
that this is a two-way street: Suppose that (by hook or by crook) we
have been provided with the RGE's but have somehow forgotten how to
calculate the effective potential. Then the RGE's can be used to {\em
renormalization group improve} the {\em classical} potential, and this
improved potential can then be used to reconstruct the one-loop
effective potential up to finite renormalizations. 

It is important to realize that there are calculational techniques, we
shall present one such later in this article, that provide the
(one-loop) RGE's without calculating the (one-loop) effective
potential. The reconstruction technique we are about to present is
then the fastest way of deriving the (one-loop) effective action.

We start with the original classical potential 
\begin{eqnarray} 
V[\phi_0] =  
{1 \over 2} m^2 \phi_0^2 + {\lambda \over {4!}} \phi_0^4. 
\end{eqnarray} 
The improvement of this classical potential consists in substituting
all the bare parameters (mass, coupling constant, wavefunction
normalization) by their renormalized running
forms, that is
\begin{eqnarray} 
m &\rightarrow& m_{\mathrm{imp}}= m(\mu), 
\\ 
\lambda &\rightarrow& \lambda_{\mathrm{imp}}= \lambda(\mu), 
\\ 
\phi &\rightarrow& \phi_{\mathrm{imp}}= \phi(\mu) = Z^{+1/2}(\mu) \phi, 
\; \; \; 
\end{eqnarray} 
to obtain 
\begin{eqnarray} 
V_{\mathrm{imp}}[\phi_0]  
&=&  {1 \over 2} m^2 (\mu) 
\phi^2_{\mathrm{imp}}  + {{\lambda (\mu)} \over {4!}} \phi^4_{\mathrm{imp}}  
\nonumber \\ 
&=& {1 \over 2} m_0^2  \; 
(\mu/\mu_0)^{{\hbar\lambda_0} \over{16 \pi^2} } \;
\exp(O(\hbar^2)) \phi_0^2 
+ \frac{\lambda_0}{4!} {\left( 
1 - {{3\hbar} \over{16 \pi^2} } \lambda_0 \ln{\mu \over \mu_0} 
 + 0(\hbar^2) \right) }^{-1} \phi_0^4. 
\end{eqnarray} 
This procedure yields an improved potential with running,
scale-dependent parameters. To one-loop we can carry out a leading-log
expansion:
\begin{eqnarray} 
V_{\mathrm{imp}}[\phi_0]
&=& {1 \over 2} m^2_0 
 {\left( 
1 +{{\hbar} \over{16 \pi^2} } \lambda_0 \ln{\mu \over \mu_0} 
\right)} 
\phi^2 
+  {\lambda_0 \over {4!}} \phi^4 
\left(1 + \frac{3\hbar}{16\pi^2} \lambda_0 
\ln\frac{\mu}{\mu_0}\right) 
+ O(\hbar^2)
\\
&=&
V[\phi_0, \mu_0] +
 {{\hbar} \over{64 \pi^2} }  
\left\{ 
{\cal M}_0^4(\mu_0) - m_0^4(\mu_0)
\right\} \ln{\mu^2 \over \mu_0^2} 
+ O(\hbar^2). 
\end{eqnarray} 
We now show that starting with the classical {\em bare} potential and
{\em improving} it, as described above, we can deduce the same results
as obtained by calculating the {\em effective } potential directly
from the functional integral (or by means of the Feynman rules), at
least up to finite renormalizations.  The key step is to realize
that we know on very general grounds (renormalizability) that the
effective potential takes the form
\begin{equation}
{\cal V}[\phi_0] = 
V[\phi_0,m(\mu),\lambda(\mu)] + 
\hbar X[\phi_0,m(\mu),\lambda(\mu),\mu] +
O(\hbar^2).
\end{equation}
We wish to see what we can deduce about the function
$X[\phi_0,m(\mu),\lambda(\mu),\mu]$. From the improved potential
discussed above we know we can write
\begin{equation}
{\cal V}[\phi_0] = 
V[\phi_0,m(\mu_0),\lambda(\mu_0)] + 
{{\hbar} \over{64 \pi^2} } 
\left\{ 
{\cal M}_0^4(\mu_0) - m_0^4
\right\} \ln{\mu^2 \over \mu_0^2}  +
\hbar X[\phi_0,m(\mu),\lambda(\mu),\mu] +
O(\hbar^2).
\end{equation}
Since the above is (despite naive appearances), {\em independent of
$\mu$}, we deduce that there must exist functions
$X_1[\phi_0,m(\mu),\lambda(\mu)]$ and
$X_2[\phi_0,m(\mu),\lambda(\mu)]$, which can no longer {\em explicitly}
depend on $\mu$, such that
\begin{equation}
X[\phi_0,m(\mu),\lambda(\mu),\mu] = 
 {{\hbar} \over{64 \pi^2} } 
\left\{ 
{\cal M}_0^4(\mu) - m^4
\right\} \ln{X_1[\phi_0,m(\mu),\lambda(\mu)] \over \mu^2} 
+ X_2[\phi_0,m(\mu),\lambda(\mu)]
+ O(\hbar).
\end{equation}
Reassembling this shows us that
\begin{equation}
{\cal V}[\phi_0] = 
V[\phi_0,m(\mu),\lambda(\mu)] + 
{{\hbar} \over{64 \pi^2} } 
\left\{ 
{\cal M}_0^4(\mu) - m^4
\right\} 
\ln{X_1[\phi_0,m(\mu),\lambda(\mu)] \over \mu^2} 
+
\hbar X_2[\phi_0,m(\mu),\lambda(\mu)] +
O(\hbar^2).
\end{equation}
We now see the beginnings of similarity with the explicit one-loop
effective potentials previously calculated. We shall complete the job
by using dimensional analysis plus the existence of appropriate limits
to constrain the functions $X_1$ and $X_2$.

For clarity of exposition it is easiest to first deal with the
massless case when the above simplifies to
\begin{equation}
{\cal V}[\phi_0] = 
V[\phi_0,\lambda(\mu)] + 
{{\hbar} \over{64 \pi^2} } 
\left\{ 
{\cal M}_0^4(\mu)
\right\} 
\ln{X_1[\phi_0,\lambda(\mu)] \over \mu^2} 
+
\hbar X_2[\phi_0,\lambda(\mu)] +
O(\hbar^2).
\end{equation}
But then by dimensional analysis
\begin{eqnarray}
X_1[\phi_0,\lambda(\mu)] &=& \alpha_1(\lambda(\mu)) \; \phi^2_0,
\\
X_2[\phi_0,\lambda(\mu)] &=& \alpha_2(\lambda(\mu)) \; \phi^4_0,
\end{eqnarray}
with the $\alpha$'s being dimensionless functions of the dimensionless
variable $\lambda$. This tells us that there exists an $\alpha_3$ such that
\begin{equation}
{\cal V}[\phi_0] = 
V[\phi_0,\lambda(\mu)] + 
{{\hbar} \over{64 \pi^2} } 
\left\{ 
{\lambda(\mu)^2 \phi_0^4(\mu) \over 4}
\right\} 
\ln{\phi_0^2 \over \mu^2} 
+
\hbar  \alpha_3(\lambda(\mu)) \; \phi_0^4 +
O(\hbar^2).
\end{equation}
Comparing this with equation (\ref{E:effective-potential:massless}) we
see that we have recovered the one-loop massless effective potential
up to an unknown term proportional to $\phi_0^4$. (That is, up to a
finite renormalization). This is of course exactly what we should
expect: the RGE's (and therefore the improved potential) are sensitive
only to the divergent terms in the renormalization, and so working
backwards from the RGE's we cannot possibly recover the terms
depending on finite renormalizations.

Repeating this procedure for the massive theory is trickier and
more tedious, but the basic ideas and results remain the same. Step
back to the general result
\begin{equation}
{\cal V}[\phi_0] = 
V[\phi_0,m(\mu),\lambda(\mu)] + 
{{\hbar} \over{64 \pi^2} } 
\left\{ 
{\cal M}_0^4(\mu) - m^4
\right\} 
\ln{X_1[\phi_0,m(\mu),\lambda(\mu)] \over \mu^2} 
+
\hbar X_2[\phi_0,m(\mu),\lambda(\mu)] +
O(\hbar^2).
\end{equation} 
Then there exists a function $X_3[\phi_0,m(\mu),\lambda(\mu)]$ such
that
\begin{equation}
{\cal V}[\phi_0] = 
V[\phi_0,m(\mu),\lambda(\mu)] + 
{{\hbar} \over{64 \pi^2} } 
\left\{ 
{\cal M}_0^4(\mu) 
\ln{ {\cal M}_0^2(\mu) \over \mu^2}
- m^4(\mu) \ln { m^2(\mu) \over \mu^2} 
\right\} 
+
\hbar X_3[\phi_0,m(\mu),\lambda(\mu)] +
O(\hbar^2).
\end{equation} 
{From} the massless case we have just investigated we know that there
exists a dimensionless function $\alpha_4$ such that
\begin{equation}
\lim_{m\to0}  X_3[\phi_0,m(\mu),\lambda(\mu)] = 
\alpha_4(\lambda(\mu)) \; \phi_0^4.
\end{equation}
But we also know (because we constructed the effective action to be
zero at zero field) that
\begin{equation}
\lim_{\phi_0\to0}  X_3[\phi_0,m(\mu),\lambda(\mu)] = 0.
\end{equation}
This is enough to tell us that there exists a dimensionless function
$\alpha_5(\lambda(\mu))$ such that
\begin{equation}
X_3[\phi_0,m(\mu),\lambda(\mu)] = 
\alpha_4(\lambda(\mu)) \; \phi_0^4 + 
\alpha_5(\lambda(\mu)) \; m^2(\mu) \; \phi_0^2.
\end{equation}
The last step is to assemble everything to see that
\begin{eqnarray}
{\cal V}[\phi_0] 
&=& 
V[\phi_0,m(\mu),\lambda(\mu)] + 
{{\hbar} \over{64 \pi^2} } 
\left\{ 
{\cal M}_0^4(\mu) 
\ln{ {\cal M}_0^2(\mu) \over \mu^2}
- m^4(\mu) \ln { m^2(\mu) \over \mu^2} 
\right\}
\nonumber
\\
&&
\qquad 
+
\hbar \left\{
\alpha_4(\lambda(\mu)) \; \phi_0^4 + 
\alpha_5(\lambda(\mu)) \; m^2(\mu) \; \phi_0^2
\right\}
+ O(\hbar^2).
\end{eqnarray} 
This completes the task we set out to perform: we have used the
one-loop improved classical potential to  reconstruct the
one-loop effective potential up to finite renormalizations.

\section{Inhomogeneous fields: Regularization} 
 
The simple momentum cutoff technique used above works well if we are
only interested in computing the effective potential, for which the
background field is taken to be homogeneous (and/or whenever
wavefunction renormalization is not an issue). For the more general
case of the full effective action where wavefunction renormalization
is important, a different method for regulating and computing the
effective action is required (or at least, preferable).  Improvement
of the complete effective action and the easy extraction of the
associated RGE's is as simple as in the case of the effective
potential.  The following is an example of one such method based on
Schwinger's proper time representation of the effective
action. Complete background details may be found in the book by
Zinn--Justin~\cite{Zinn-Justin}, (Appendix to Chapter $8$ in second
edition, Appendix to Chapter $9$ in third edition), in which the
divergences of the effective action (up to one-loop) are calculated
explicitly.
 
For the most general scalar field action we can write 
\begin{eqnarray} 
S[\phi]= \int \d^n x \left\{ 
{1 \over 2} \partial_\mu \phi(x)\,  \partial^\mu \phi(x) + V[\phi]
\right\}, 
\end{eqnarray} 
with 
\begin{eqnarray} 
V[\phi]&=& 
\sum_{j=0}^{N} {\lambda_j \over {j!}} \phi^j. 
\end{eqnarray} 
The operator $S_2$ takes the form 
\begin{eqnarray} 
S_2(x_1,x_2;\phi)= 
[- \partial_\mu \partial^\mu + {V}'' [\phi (x_1)]] 
\delta^n (x_1,x_2), 
\end{eqnarray} 
and 
\begin{eqnarray} 
S_2(x_1,x_2;\phi=0)= 
[- \partial_\mu \partial^\mu + m^2 ] 
\delta^n (x_1,x_2), 
\end{eqnarray} 
with the special notation that $m^2 = \lambda_2  = V''[\phi=0]$.
 
We now make use of the result derived by
Zinn--Justin~\cite{Zinn-Justin} for the
divergent term in the one-loop correction to the effective action
(using Schwinger's proper time regularization):
\begin{eqnarray}\label{E:Schwinger} 
\Gamma_1^\epsilon [\phi]&=& {1 \over 2} {1 \over {{(4 \pi)}^{n/2}}} 
\left\{ 
- {{\epsilon^{1-n/2}} \over {1-n/2}} \int \d^n x \; [{V}''(x) - m^2] 
+{1 \over 2} {{\epsilon^{2-n/2}} \over {2-n/2}} \int \d^n x \;  
[{(V'')}^2(x) - m^4] 
\right. 
\\ \nonumber 
&-& 
\left. 
{1 \over 6} {{\epsilon^{3-n/2}} \over {3-n/2}} \int \d^n x \;  
\left[ {(V'')}^3(x) - m^6 + {1 \over 2}{\partial_\mu {V}''(x)} 
\; {\partial^\mu {V}''(x)} 
\right] 
\right\} + \dots. 
\end{eqnarray} 
See (A8.15) in the Appendix to Chapter $8$ of the second edition, or
(A9.15) in the Appendix to Chapter $9$ of the third edition.  Here a
comparison with the usual cutoff regulator can be made by replacing
$\epsilon \to \Lambda^{-2}$. Additionally, whenever $n$ is an even
integer $\epsilon^0/0$ is to be understood as $\ln(1/\epsilon)$, and
for comparison with the cutoff regulator should be replaced by
$\ln(\Lambda^2/\mu^2)$. This expression gives all the divergences
for $n \leq 6$.
 
We are now ready to apply this result to a particular model. We shall 
first repeat the calculation of the RGE's for $\lambda \phi^4$ theory, 
in order to compare with the results obtained using the momentum 
cutoff method in the previous section. Notice that the relevance of 
the expression (\ref{E:Schwinger}) is the fact that we can carry out the 
renormalization at the level of the effective action, and not merely 
the effective potential, therefore we can deal with the kinetic term 
and with wavefunction renormalizations, as can be appreciated by the 
appearance of the gradient terms. 

\section{Effective action for $\lambda(\phi^4)_4$}
 
The divergent terms of the effective action for the $\lambda \phi^4$ 
theory in four dimensions are 
\begin{eqnarray} 
\Gamma_1^\Lambda [\phi]
&=&  
{1 \over {{64 \pi^2}}} 
\left\{ 
{{\Lambda^{2}}}  \int \d^4 x \; \lambda \phi^2 (x) 
- \ln {\Lambda^2 \over m^2} \int \d^4 x \;  
\left[ {\lambda^2 \over 4} \phi^4 (x) + \lambda m^2 \phi^2 (x) 
 \right] 
\right\}
\\
&=&
{1 \over {{64 \pi^2}}} 
\left\{ 
{{\Lambda^{2}}}  \int \d^4 x \; \lambda \phi^2 (x) 
- \ln {\Lambda^2 \over m^2} \int \d^4 x \;  
\left[ {\cal M}^4(\phi(x)) - m^4  \right] 
\right\}. 
\end{eqnarray} 
Here ${\cal M}^4(\phi(x)) = m^2 + (\lambda/2)\phi(x)^2$ is the
position dependent generalization of the ${\cal M}(\phi_0)$ defined in
the previous analysis. For a  constant field of course, this reproduces
the divergent pieces exhibited in equation
(\ref{E:effective-potential:bare}).  The fact that there is no
divergence containing a gradient term is the justification for our
statement that the one-loop wavefunction renormalization for
$\lambda(\phi^4)_4$ vanishes. By this derivation the result is clearly
much more general: for {\em any} scalar field theory in dimensions
five or less (renormalizable or not) the one-loop wavefunction
renormalization vanishes. (In particular for $\lambda(\phi^6)_3$ and
$\lambda(\phi^n)_2$ at one loop all the divergences can be collected
in the effective potential.) Putting all this aside for now, we focus
attention on $\lambda(\phi^4)_4$ and check that the RGE's are
completely independent of the regularization scheme.

In the one-loop approximation 
\begin{eqnarray} 
\Gamma[\phi]&=& S[\phi] - S[0] + \hbar \Gamma_1[\phi] + O(\hbar^2),
\end{eqnarray}
with
\begin{eqnarray} 
\Gamma_1[\phi] &=&
\Gamma_1^\Lambda[\phi]+ \Gamma_1^{\mathrm{finite}}[\phi]. 
\end{eqnarray} 
Since this is a renormalizable theory, we can hope to absorb the
divergences into redefinitions of the parameters. For instance we can
write
\begin{eqnarray} 
S[\phi]&=&   S[\phi, \mu] 
- {\hbar \over {{64 \pi^2}}} 
\left\{ 
 {{\Lambda^{2}}}  \int \d^4 x \; \lambda \phi^2 (x) 
- \ln {\Lambda^2 \over \mu^2} \int \d^4 x \;  
\left[ {\cal M}^4(\phi(x)) - m^4  \right] 
+ O(\hbar^2) 
\right\}, 
\end{eqnarray} 
so that the effective action may be rewritten as 
\begin{eqnarray} 
\Gamma[\phi]&=& S[\phi, \mu]  
+ {1 \over {{32 \pi^2}}} \ln {m \over \mu}  
\int \d^4 x \;  
\left[ {\lambda^2 \over 4} \phi^4 (x) + \lambda m^2 \phi^2 (x) 
 \right] 
+ 
\Gamma_1^{\mathrm{finite}}[\phi] . 
\end{eqnarray} 
This is not quite the same decomposition into renormalized parameters
and counterterms as made previously [equation (\ref{E:counterterms})],
differing from that prescription by a finite renormalization. {\em
This does not matter and does not change any physics}. Also note that
$\Gamma_1^{\mathrm{finite}}[\phi]$ contains both non-gradient terms
(equivalent to the previously derived effective potential up to finite
renormalizations) and gradient terms that were absent in the effective
potential calculation. (These gradient terms start out at
$O(\hbar\,\phi^2\,(\partial\phi)^2)$, and contain arbitrarily high
powers of both the field and gradients of the field, but the
coefficients of these terms are both finite and [in principle]
calculable, which is what makes renormalizable theories predictively
useful.)

We are now ready to derive the renormalization group equations, but 
before doing so, and for the sake of clarity, we give the expression 
for the scale-dependent classical action, that is the classical action 
with scale-dependent parameters. 
\begin{eqnarray} 
S[\phi, \mu]= \int \d^4 x \;  
{1 \over 2}  
Z(\mu) \;
\partial_\mu \phi(x) \; \partial^\mu \phi(x) + {1 \over 2}  
Z(\mu) \; m^2 (\mu) 
\phi^2(x) + {{\lambda (\mu)} \over {4!}}  Z^{2}(\mu) \; \phi^4(x). 
\end{eqnarray} 
As we have already seen, the renormalization group equations follow
from
\begin{eqnarray} 
\mu \; {{\d  \Gamma[\phi]} \over {\d\mu}} = 0 = 
\mu \; {{\d  S[\phi, \mu]} \over {\d\mu}} + \hbar \mu  
{{\d  \Gamma_1[\phi]} \over {\d\mu}} + O(\hbar^2), 
\end{eqnarray} 
which yields 
\begin{eqnarray} 
&+&{\mu \over 2} {{\d Z(\mu)} \over {\d \mu}} 
({\partial_\nu \phi} \, {\partial^\nu \phi}) 
\nonumber\\ 
&+& 
{\mu \over 2} Z(\mu) {{\d m^2(\mu)} \over {\d \mu}} {\phi}^2 
+{\mu \over 2} m^2(\mu)  {{\d Z(\mu)} \over {\d \mu}} {\phi}^2 
\nonumber\\ 
&+& 
{\mu \over {4!}}  
Z^{2}(\mu) {{\d \lambda (\mu)} \over {\d \mu}} {\phi}^4 
+{2 \mu} {{\lambda (\mu)} \over {4!}}  
Z(\mu) {{\d Z(\mu)} \over {\d \mu}} {\phi}^4 
\nonumber\\
&&\qquad =
{\hbar \over {{32 \pi^2}}}  
\left( {\lambda^2 \over 4} \phi^4  + \lambda m^2 \phi^2  \right) 
+ O(\hbar^2). 
\end{eqnarray} 
We now proceed to identify terms in the left and right hand sides 
of the previous equation:
\begin{eqnarray} 
{\mu \over 2} {{\d Z} \over {\d \mu}} 
({\partial_\nu \phi} \, {\partial^\nu \phi})
&=& 0 + O(\hbar^2)
\\  
{\mu \over 2} Z {{\d m^2} \over {\d \mu}} {\phi}^2 
+{\mu \over 2} m^2  {{\d Z} \over {\d \mu}} {\phi}^2 
&=&
{\hbar \over {{32 \pi^2}}}  \lambda m^2 \phi^2 
+O(\hbar^2)
\\ 
{\mu \over {4!}}  
Z^{2} {{\d \lambda} \over {\d \mu}} {\phi}^4 
+{2 \mu} {\lambda \over {4!}}  
Z {{\d Z} \over {\d \mu}} {\phi}^4
&=&
{\hbar \over {{32 \pi^2}}} {\lambda^2 \over 4} \phi^4 
+O(\hbar^2) 
. 
\end{eqnarray} 
We can conclude then 
\begin{eqnarray} 
{{\d Z} \over {\d \mu}}  
&=& 0 +O(\hbar^2)
\\  
{\mu \over 2} Z {{\d m^2} \over {\d \mu}} {\phi}^2 
&=&
{\hbar \over {{32 \pi^2}}} \lambda m^2 \phi^2 
+O(\hbar^2)
\\ 
{\mu \over {4!}}  
Z^{2}{{\d \lambda} \over {\d \mu}} {\phi}^4 
&=&
{\hbar \over {{32 \pi^2}}}  
 {\lambda^2 \over 4} \phi^4 
+O(\hbar^2). 
\end{eqnarray} 
{From} the first equation, and not to change the normalization from that
of the bare fields, we can choose $Z(\mu) = 1 + O(\hbar^2)$, which
yields the following two renormalization group equations for
$m^2(\mu)$ and $\lambda (\mu)$
\begin{eqnarray} 
{\mu} {{\d m^2} \over {\d\mu}} 
&=&
{\hbar \over {16 \pi^2}}  m^2 \lambda  
+O(\hbar^2) 
\\  
\mu {{\d \lambda} \over {\d\mu}} 
&=&
{{3 \hbar} \over {16 \pi^2}}  {\lambda^2} 
+O(\hbar^2), 
\end{eqnarray} 
which can be seen to be the same as the ones
(\ref{E:rge-1}--\ref{E:rge-2}) derived using the effective potential
approach with a momentum cutoff regularization prescription.  [This
verifies (at one loop) the scheme independence of the RGE's.] Notice
that here we have by no means made use of the assumption that the
field was homogeneous. Also, although the wavefunction renormalization
is $Z(\mu) = 1 + O(\hbar^2)$ to this order, the proper-time formalism
is well-suited to handling divergences in the kinetic energy term of
the effective action.  To see how this works, we will subsequently
apply this technique to the $\lambda \phi^3$ theory in $6$ dimensions, 
where there is a one-loop wavefunction renormalization.
 
Before leaving the $\lambda(\phi^4)_4$ theory, we will explicitly
point out that with the current technique we have just computed the
one-loop RGE's without calculating the one-loop effective action (or
even the one-loop effective potential). This situation is made to
order for our reconstruction technique, now using the renormalization
group improved classical {\em action}.  In the same
way that we improved the classical potential we now consider
\begin{eqnarray} 
S[\phi(x)] = \int \d^4x \left\{
{1 \over 2} \, \partial_\mu \phi \; \partial^\mu \phi 
+ {1 \over 2} \, m^2 \;
\phi^2 + {\lambda \over {4!}} \phi^2
\right\}. 
\end{eqnarray} 
We {\em improve} this classical action by substituting all the
parameters by the improved running parameters, that is
\begin{eqnarray} 
m &\rightarrow& m_{\mathrm{imp}}= m(\mu) 
\\ 
\lambda &\rightarrow& \lambda_{\mathrm{imp}}= \lambda(\mu) 
\\ 
\phi &\rightarrow& \phi_{\mathrm{imp}}= \phi(\mu)=Z^{1/2}(\mu) \; \phi, 
\end{eqnarray} 
to obtain 
\begin{eqnarray} 
S_{\mathrm{imp}}
&=& \int \d^4x \left\{
{1 \over 2} 
\partial_\mu \phi_{\mathrm{imp}}  \, 
\partial^\mu \phi_{\mathrm{imp}}  
+ {1 \over 2} m^2 (\mu) 
\phi^2_{\mathrm{imp}}  
+ {{\lambda (\mu)} \over {4!}} \phi^4_{\mathrm{imp}} 
\right\} 
\\ \nonumber 
&=& \int \d^4x \left\{
{1 \over 2} Z(\mu) \partial_\mu \phi \, \partial^\mu \phi 
+ {1 \over 2} m^2 (\mu)  Z(\mu)  \phi^2  
+ {{\lambda (\mu)} \over {4!}}  Z^{2}(\mu)  \phi^4
\right\}. 
\end{eqnarray} 
We now follow exactly the same procedure as for the effective
potential, expanding the above in leading logarithms to deduce the
$\mu$ dependence of the improved action, which then gives information
about the $\mu$ dependence of the one-loop contribution to the
effective action. Simply repeating the previous steps now yields
\begin{eqnarray}
\Gamma[\phi(x)] 
&=& 
S[\phi(x),m(\mu),\lambda(\mu)] + 
\int \d^4x \Bigg[
{{\hbar} \over{64 \pi^2} } 
\left\{ 
{\cal M}^4(\phi(x),\mu) 
\ln{ {\cal M}^2(\phi(x),\mu) \over \mu^2}
- m^4(\mu) \ln { m^2(\mu) \over \mu^2} 
\right\}
\nonumber
\\
&&
\qquad 
+
\hbar \left\{
\alpha_4(\lambda(\mu)) \; \phi(x)^4 + 
\alpha_5(\lambda(\mu)) \; m^2(\mu) \; \phi(x)^2
\right\}
+ O(\hbar \phi^2 (\partial\phi)^2)  + O(\hbar(\partial\phi)^4) \Bigg]
+ O(\hbar^2).
\end{eqnarray} 
The novelty here is that the field $\phi(x)$ is allowed to be position
dependent. We have the same finite renormalization ambiguities as in
the constant field (effective potential) case, but there are now
additional unknowns, even at one loop, coming from higher orders in
the gradient expansion. The coefficients of the $O(\hbar \phi^2
(\partial\phi)^2)$ and $O(\hbar(\partial\phi)^4)$ terms are finite and
in principle calculable, and so are not accessible via RGE
techniques---we will have to resort to a Feynman diagram or related
type of calculation to extract the actual coefficient. For
completeness, we mention that Itzykson and Zuber report the $O(\hbar
\phi^2 (\partial\phi)^2)$ term to be
\begin{equation} 
{\hbar \lambda^2\over192\pi^2} 
{\phi^2 (\partial\phi)^2\over m^2+ {\lambda\over2} \phi^2},
\end{equation} 
(see p. 455, eq. (9-130) of \cite{Itzykson-Zuber}.) 

The significance of the results presented so far lies not in that we
have complete information regarding the effective potential (which we
do not), but rather in that after this long (because we have been very
explicit) build-up, we have a method for very rapidly extracting the
one-loop effective action with a minimum of actual calculation.

We shall now apply this technique to $\lambda(\phi^3)_6$ (where we
have to worry about wavefunction renormalization even at one loop), to
$\lambda(\phi^6)_3$ (where everything is particularly simple at one
loop), and to $\lambda(\phi^n)_2$ (where in 2 dimensions almost
anything is renormalizable).

\section{Renormalization group equations for $\lambda (\phi^3)_6$ theory} 

Let us consider now $\lambda (\phi^3)_6$. The bare action is
\begin{eqnarray} 
S[\phi]= \int \d^6 x \;  
\left\{
{1 \over 2} \partial_\mu \phi(x) \, \partial^\mu \phi(x) + 
{\lambda_1 \over {1!}} \phi (x) + 
{1 \over 2!} m^2 \phi^2(x) + 
{\lambda_3 \over {3!}} \phi^3 (x)
\right\}. 
\end{eqnarray} 
Therefore
\begin{eqnarray} 
{V}'' &=& \lambda_2 + \lambda_3 \phi 
\define m^2 + \lambda_3 \phi. 
\end{eqnarray} 
The divergent terms of the effective action for the $\lambda \phi^3$ 
theory in $6$ dimensions are 
\begin{eqnarray} 
\Gamma_1^\Lambda [\phi]&=&  
 {1 \over {{128 \pi^3}}} 
\left\{ 
 {{\Lambda^{4}} \over 2}  \int \d^6 x \; \lambda_3 \phi (x) 
-  {{\Lambda^{2}} \over 2}  \int \d^6 x \; [\lambda_3^2 \phi^2 (x) 
+ 2 m^2 \lambda_3 \phi (x) ] 
\right. 
\\ \nonumber 
&+& 
\left. 
 {1 \over 3} \ln {\Lambda \over m} \int \d^6 x \;  
\left[ {\lambda_3^3} \phi^3 (x) + 3 \lambda_3^2 m^2 \phi^2 (x) 
+ 3 m^4 \lambda_3 \phi(x) + 
{\lambda_3^2 \over 2} \partial_\mu \phi (x) \partial^\mu \phi (x) 
 \right] 
\right\}. 
\end{eqnarray} 
At the one-loop approximation 
\begin{eqnarray} 
\Gamma[\phi]&=& S[\phi] -S[0] + \hbar\Gamma_1[\phi] +O(\hbar^2),
\end{eqnarray}
with
\begin{eqnarray}
\Gamma_1[\phi]&=& \Gamma_1^\Lambda[\phi]+ \Gamma_1^{\mathrm{finite}}[\phi]. 
\end{eqnarray} 
A minor caution: $S[0]$ means $S[\phi[J=0]]$, which is not zero for an
asymmetric theory like this. Fortunately, $S[0]$ is by definition a
field-independent offset to the effective action, which when written
in terms of bare parameters is manifestly independent of the
renormalization scale $\mu$. It therefore does not contribute to the
RGE's (at any number of loops), and can be quietly ignored.

This is a renormalizable theory, and therefore we can split the bare
parameters into renormalized parameters and counterterms and write
\begin{eqnarray} 
S[\phi]
&=&   
S[\phi, \mu] 
- {\hbar \over {{128 \pi^3}}} 
\left\{ 
 {{\Lambda^{4}} \over 2}  \int \d^6 x \; \lambda_3 \phi (x) 
-  {{\Lambda^{2}} \over 2}  \int \d^6 x \; [\lambda_3^2 \phi^2 (x) 
+ 2 m^2 \lambda_3 \phi (x) ] 
\right. 
\\ \nonumber 
&+& 
\left. 
 {1 \over 3} \ln {\Lambda \over \mu} \int \d^6 x \;  
\left[ {\lambda_3^3} \phi^3 (x) + 3 \lambda_3^2 m^2 \phi^2 (x) 
+ 3 m^4 \lambda_3 \phi(x) + 
{\lambda_3^2 \over 2} \partial_\mu \phi (x) \partial^\mu \phi (x) 
 \right] 
\right\} +O(\hbar^2), 
\end{eqnarray} 
so that the effective action can be rewritten as 
\begin{eqnarray} 
\Gamma[\phi]&=& S[\phi, \mu] - S[0]
+ {\hbar \over {{384 \pi^3}}} 
 \ln {\mu \over m} \int \d^6 x \;  
\left[ {\lambda_3^3} \phi^3 (x) + 3 \lambda_3^2 m^2 \phi^2 (x) 
+ 3 m^4 \lambda_3 \phi(x) + 
{\lambda_3^2 \over 2} \partial_\mu \phi (x) \partial^\mu \phi (x) 
 \right] 
\nonumber\\
&& \qquad
+\hbar \Gamma_1^{\mathrm{finite}}[\phi] 
+O(\hbar^2).
\end{eqnarray} 
We are now ready to derive the corresponding renormalization group
equations, but before doing so, and for the sake of clarity, we give
the expression for the scale-dependent classical action, that is, the
classical action written in terms of scale-dependent parameters. (This
expression is exact, though as a practical matter calculations are
typically carried out at some fixed order in the loop expansion.)
\begin{eqnarray} 
S[\phi, \mu]= \int \d^6 x \;  
{1 \over 2}  
Z(\mu) \;
\partial_\mu \phi(x) \; \partial^\mu \phi(x) 
+ {{\lambda_1 (\mu)} \over {1!}}  Z^{1/2}(\mu) \; \phi(x) 
+ {m^2(\mu) \over 2!}  Z(\mu) \; \phi^2(x) 
+ {{\lambda_3 (\mu)} \over {3!}}  Z^{3/2}(\mu) \; \phi^3(x). 
\end{eqnarray} 
Again, the renormalization group equations follow immediately from
\begin{eqnarray} 
\mu \; {{\d \; \Gamma[\phi]} \over {\d\mu}} 
= 0 = 
\mu \; {{\d \; S[\phi, \mu]} \over {\d\mu}} + \hbar \mu  
{{\d \; \Gamma_1[\phi]} \over {\d\mu}} 
+O(\hbar^2), 
\end{eqnarray} 
which yields 
\begin{eqnarray}
&+&{\mu \over 2} {{\d Z(\mu)} \over {\d \mu}} 
({\partial_\nu \phi} \, {\partial^\nu \phi} )
\\ \nonumber 
&+& 
{\mu} Z^{1/2}(\mu) {{\d \lambda_1(\mu)} \over {\d \mu}} {\phi} 
+{\mu \over 2} \lambda_1(\mu)  Z^{-1/2}(\mu) 
{{\d Z(\mu)} \over {\d \mu}} {\phi} 
\\ \nonumber 
&+& 
{\mu \over 2} Z(\mu) {{\d m^2(\mu)} \over {\d \mu}} {\phi}^2 
+{\mu \over 2} m^2(\mu) {{\d Z(\mu)} \over {\d \mu}} {\phi}^2 
\\ \nonumber 
&+& 
{\mu \over {3!}}  
Z^{3/2}(\mu) {{\d \lambda_3 (\mu)} \over {\d \mu}} {\phi}^3 
+{{3 \mu} \over 2} {{\lambda_3 (\mu)} \over {3!}}  
Z^{1/2}(\mu) {{\d Z(\mu)} \over {\d \mu}} {\phi}^3
\\ \nonumber
&& \qquad = 
- {\hbar \over {{384 \pi^3}}} 
  \left[ {\lambda_3^3} \phi^3 + 3 \lambda_3^2 m^2 \phi^2  
+ 3 m^4 \lambda_3 \phi + 
{\lambda_3^2 \over 2} \partial_\mu \phi  \partial^\mu \phi  
 \right] 
+O(\hbar^2).
\end{eqnarray} 
Matching like powers of $\phi$ on both sides of this equation, as we
did previously, we deduce the following set of (1-loop)
renormalization group equations:
\begin{eqnarray} 
\mu {{\d Z} \over {\d \mu}}  
&=&
- {\hbar \over {384 \pi^3}} \lambda_3^2 
+O(\hbar^2)
\\  
{\mu} Z {{\d \lambda_1} \over {\d \mu}}  
+ {\mu \over 2} \lambda_1   {{\d Z} \over {\d \mu}}  
&=&
- {\hbar \over {{128 \pi^3}}}  \lambda_3 m^4 Z^{1/2} 
+O(\hbar^2)
\\  
{\mu} Z {{\d m^2} \over {\d \mu}} + {\mu} m^2  {{\d Z} \over {\d \mu}}  
&=&
- {\hbar \over {{64 \pi^3}}} \lambda_3^2 m^2 
+O(\hbar^2)
\\ 
{\mu}  Z{{\d \lambda_3} \over {\d \mu}} 
+ {{3 \mu} \over 2} \lambda_3 {{\d Z} \over {\d \mu}}
&=&
-{\hbar \over {{64 \pi^3}}} \lambda_3^3 Z^{-1/2} 
+O(\hbar^2).
\end{eqnarray} 
We now define the following functions 
\begin{eqnarray} 
\beta_1 \left( \lambda_1, \lambda_3, {m \over \mu} \right) 
&\define& 
\mu {{\partial \lambda_1}\over {\partial \mu}} 
\\ 
\beta_3 \left( \lambda_1, \lambda_3, {m \over \mu} \right) 
&\define& 
\mu {{\partial \lambda_3}\over {\partial \mu}} 
\\ 
\gamma_{\mathrm{m}} \left( \lambda_1, \lambda_3, {m \over \mu} \right) 
&\define& 
\mu m^{-2} {{\partial m^2}\over {\partial \mu}} 
\\ 
\gamma_{\mathrm{d}} \left( \lambda_1, \lambda_3, {m \over \mu} \right) 
&\define&  
\mu {{\partial \ln \; Z}\over {\partial \mu}} , 
\end{eqnarray} 
to obtain 
\begin{eqnarray} 
\beta_1 \left( \lambda_1, \lambda_3, {m \over \mu} \right) 
&=& 
+ \hbar\left\{ {{\lambda_1 \lambda_3^2} \over {768 \pi^3}} 
  - {{\lambda_3 m^4} \over {128 \pi^3}} \right\}
+ O(\hbar^2)
\\ 
\beta_3 \left( \lambda_1, \lambda_3, {m \over \mu} \right) 
&=&
- {{3 \hbar \lambda_3^3} \over {256 \pi^3}} 
+O(\hbar^2)
\\ 
\gamma_{\mathrm{m}} \left( \lambda_1, \lambda_3, {m \over \mu} \right) 
&=&  
-{{5 \hbar\lambda_3^2} \over {384 \pi^3}} 
+O(\hbar^2)
\\ 
\gamma_{\mathrm{d}} \left( \lambda_1, \lambda_3, {m \over \mu} \right) 
&=&  
-{\hbar {\lambda_3^2} \over {384 \pi^3}} 
+ O(\hbar^2). 
\end{eqnarray} 
These results can be checked (for example) against chapter 7, section
3, of Collins~\cite{Collins}.  (We again warn the reader that there
are conflicting semi-standard definitions for the anomalous dimensions
$\gamma$, our choice is opposite to that for Collins~\cite{Collins},
but the same as that of Ramond~\cite{Ramond} and
Zinn--Justin~\cite{Zinn-Justin}.) We do not need to check at the level
of the primitive parameters $m^2$, $Z$ and $\lambda$, because it is more
efficient to do so at the level of the parameters $\beta_3$,
$\gamma_{\mathrm{m}}$, and $\gamma_{\mathrm{d}}$. In most treatments
the equation for $\beta_1$ is not taken into consideration.  It is
traditionally assumed that there is no term linear in $\phi$ in the
original action, and furthermore, that when renormalizing, we should
impose the condition that all the tadpole diagrams vanish: that is,
that the renormalized and scale dependent $\lambda_1(\mu)$ must
vanish. This assumption is tantamount to fine-tuning the bare
parameter $\lambda_1$, order by order in perturbation theory, to {\em
force} $\lambda_1(\mu)=0$.

A simplifying approach that does not involve fine-tuning is to go to
the massless and zero external bias limit. That is, to simultaneously
set $m=0$ and $\lambda_1=0$, this being a fixed-plane of the one-loop
RGE's. Then we can safely restrict attention to the
$\beta_3$--$\gamma_{\mathrm{d}}$ plane. (We can also easily check that
if we simultaneously set both bare parameters $\lambda_1$ and $m$
equal to zero, then renormalization effects will not generate such
parameters, at least to one loop order in six dimensions.)

\section{Improved action for the massless zero-bias $\lambda (\phi^3)_6$ 
theory} 

We start with the bare massless Lagrangian density for $\lambda
(\phi^3)_6$:
\begin{eqnarray} 
{\cal L} = {1 \over 2} \partial_\mu \phi \partial^\mu \phi 
+ {\lambda \over {3!}} \phi^3. 
\end{eqnarray} 
The {\em improvement} of this Lagrangian density consists in
substituting all the bare parameters by the renormalization group
improved ones, that is
\begin{eqnarray} 
\lambda &\rightarrow& \lambda_{\mathrm{imp}}= \lambda(\mu) 
\\ 
\phi &\rightarrow& \phi_{\mathrm{imp}}= \phi(\mu)=Z^{1/2}(\mu) \; \phi, 
\end{eqnarray} 
to obtain 
\begin{eqnarray} 
{\cal L}_{\mathrm{imp}}  
&=& 
{1 \over 2} \partial_\mu \phi_{\mathrm{imp}}  \, 
\partial^\mu \phi_{\mathrm{imp}}  
+ {{\lambda (\mu)} \over {3!}} \phi^3_{\mathrm{imp}}  
\\ \nonumber 
&=& {1 \over 2} Z(\mu) \partial_\mu \phi  \,
 \partial^\mu \phi 
+ {{\lambda (\mu)} \over {3!}}  Z^{3/2} (\mu)  \phi^3. 
\end{eqnarray} 
In this massless zero-bias case the RGE's are straightforward to
solve. We have
\begin{equation}
\lambda(\mu) = 
\lambda(\mu_0) 
\left(
1 + {3\hbar\lambda^2(\mu_0)\over128\pi^3} \ln{\mu\over\mu_0} +O(\hbar^2)
\right)^{-1/2},
\end{equation}
and
\begin{equation}
Z(\mu) = Z(\mu_0) \;
\left(\mu/\mu_0\right)^{-{\hbar\lambda^2(\mu_0)\over384\pi^3}}
\exp(O(\hbar^2)).
\end{equation}
Inserting this into the improved action and expanding to $O(\hbar^2)$
we see
\begin{eqnarray} 
{\cal L}_{\mathrm{imp}}  
&=& {1 \over 2} 
Z(\mu_0) \; \partial_\mu \phi  \; \partial^\mu \phi 
+ {{\lambda (\mu_0)} \over {3!}}  Z(\mu_0)^{3/2}  \phi^3
\nonumber\\
&& \qquad
+ \hbar \left\{
-{\lambda^2(\mu_0)\over768\pi^3}
  Z(\mu_0) \; \partial_\mu \phi  \; \partial^\mu \phi 
-{\lambda^3(\mu_0)\over 384\pi^3}  Z(\mu_0)^{3/2}  \; \phi^3
\right\} \ln{\mu\over\mu_0}
+O(\hbar^2)
\\
&=& {1 \over 2} 
\partial_\mu \phi(\mu_0)  \; \partial^\mu \phi(\mu_0) 
+ {{\lambda (\mu_0)} \over {3!}}    \phi(\mu_0)^3
\nonumber\\
&&\qquad
+ \hbar \left\{
-{\lambda^2(\mu_0)\over768\pi^3}
  \partial_\mu \phi(\mu_0)  \; \partial^\mu \phi(\mu_0) 
-{\lambda^2(\mu_0)\over 384\pi^3}   \phi(\mu_0)^3
\right\} \ln{\mu\over\mu_0}
+O(\hbar^2).
\end{eqnarray} 
As we have seen before in other examples, the explicit $\mu$
dependence in the improved action must cancel against the explicit
$\mu$ dependence in the one-loop contribution. Without repeating the
details (already presented in the $\lambda(\phi^4)_4$ calculation),
this constrains the form of the one-loop effective potential and
implies that up to finite renormalizations
\begin{eqnarray} 
{\cal L}_{\mathrm{effective}}  
&=& 
{1 \over 2} 
\partial_\mu \phi(\mu)  \; \partial^\mu \phi(\mu) 
+ {{\lambda (\mu)} \over {3!}}  \phi(\mu)^3
+ \hbar \left\{
-{\lambda(\mu)^2\over1536\pi^3}
  \partial_\mu \phi(\mu)  \, \partial^\mu \phi(\mu) 
-{\lambda(\mu_0)^3\over 768\pi^3}   \phi(\mu)^3
\right\} \ln{\phi(\mu)\over\mu^2}
\nonumber\\
&&\qquad \qquad
+O(\hbar \phi (\partial\phi)^2) +O(\hbar^2).
\end{eqnarray} 
Part of this expression can be checked against the effective
potential. (The logarithms in the effective action have to match up
with the logarithmically divergent terms in the regulated one-loop
effective potential.) But the term involving gradients that arises
from the one-loop wavefunction renormalization is completely
inaccessible via effective potential techniques.

\section{RGE's and effective action for $\lambda (\phi^6)_3$} 

The key result can be very succinctly stated: at one loop nothing runs
in 3 dimensions. Once we have seen this result once, it becomes
obvious from equation (\ref{E:Schwinger}) that this is generic to any
odd-dimensional spacetime at one loop. In any odd number of dimensions
there will be no logarithmic divergences, and hence no terms logarithmic
in the renormalization scale in the one-loop effective action. 

So let us turn to considering our simple example: symmetric
$\lambda(\phi^6)_3$. The bare action is
\begin{eqnarray} 
S[\phi]= \int \d^3 x \;  
\left\{
{1 \over 2} \partial_\mu \phi(x) \, \partial^\mu \phi(x) + 
{1 \over 2!} m^2 \phi^2(x) + 
{\lambda_4 \over {4!}} \phi^4 (x) +
{\lambda_6 \over {6!}} \phi^6 (x)
\right\}. 
\end{eqnarray} 
Therefore
\begin{eqnarray} 
{V}'' &=&  m^2 + {1\over2!}\lambda_4 \phi^2 + {1\over4!} \lambda_6 \phi^4. 
\end{eqnarray} 
The divergent terms of the effective action for the $\lambda \phi^6$ 
theory in $3$ dimensions are 
\begin{eqnarray} 
\Gamma_1^\Lambda [\phi]&=&  
 {1 \over {(4\pi)^{3/2}}} 
\left\{ 
 {{\Lambda}}  \int \d^3 x \; 
\left[ {1\over2!}\lambda_4 \phi^2 + {1\over4!} \lambda_6 \phi^4 \right]
+ O(\Lambda^{-1})
\right\}. 
\end{eqnarray} 
Note that there are no logarithmic divergences (this will play a very
important role!).  At the one-loop approximation, we again know that
\begin{eqnarray} 
\Gamma[\phi]&=& S[\phi] -S[0] + \hbar\Gamma_1[\phi] +O(\hbar^2),
\end{eqnarray}
with
\begin{eqnarray}
\Gamma_1[\phi]&=& \Gamma_1^\Lambda[\phi]+ \Gamma_1^{\mathrm{finite}}[\phi]. 
\end{eqnarray} 
This is a renormalizable theory, and therefore we can split the bare
parameters into renormalized parameters and counterterms to write
\begin{eqnarray} 
S[\phi]
&=&   
S[\phi, \mu] 
- {\hbar  \over {(4\pi)^{3/2}}} 
\left\{ 
 {{\Lambda}}  \int \d^3 x \; 
\left[ {1\over2!}\lambda_4 \phi^2 + {1\over4!} \lambda_6 \phi^4 \right]
+ O(\Lambda^{-1})
\right\} + O(\hbar^2), 
\end{eqnarray} 
so that the effective action can be rewritten as 
\begin{eqnarray} 
\Gamma[\phi]&=& S[\phi, \mu] - S[0]
+\hbar \Gamma_1^{\mathrm{finite}}[\phi] 
+O(\hbar^2).
\end{eqnarray} 
We are now ready to derive the corresponding renormalization group
equations. Since the renormalization scale does not appear in
$\Gamma_1^{\mathrm{finite}}$ above, the RGE's will be trivial to one
loop. Differentiating
\begin{eqnarray} 
\mu \; {{\d \; \Gamma[\phi]} \over {\d\mu}} 
= 0 = 
\mu \; {{\d \; S[\phi, \mu]} \over {\d\mu}} + \hbar \mu  
{{\d \; \Gamma_1[\phi]} \over {\d\mu}} 
+O(\hbar^2), 
\end{eqnarray} 
yields 
\begin{eqnarray} 
\mu {{\d Z} \over {\d \mu}}  
&=&
0
+O(\hbar^2)
\\  
{\mu} {{\d m^2} \over {\d \mu}}  
&=&
0
+O(\hbar^2)
\\  
{\mu} {{\d \lambda_4} \over {\d \mu}} 
&=&
0
+O(\hbar^2)
\\ 
{\mu}  {{\d \lambda_6} \over {\d \mu}} 
&=&
0
+O(\hbar^2).
\end{eqnarray} 
Thus all the $\beta$ functions vanish at one loop. This does {\em not}
mean that this theory is one-loop finite. It does mean that the theory
does not ``run'' at one loop, and that the renormalized coupling
constants are renormalization scale independent at this order.  The
effective potential is simply equal to the classical potential, up to
finite renormalization ambiguities. The effective action, explicitly
exhibiting finite renormalization ambiguities, is simply
\begin{eqnarray} 
{\cal L}_{\mathrm{effective}}  
&=& {1 \over 2} 
\partial_\mu \phi  \; \partial^\mu \phi +
{1 \over 2!} m^2 \; \phi^2(x) + 
{\lambda_4 \over {4!}} \; \phi^4 (x) +
{\lambda_6 \over {6!}} \; \phi^6 (x)
\nonumber\\
&&\qquad
+ \hbar \left\{ 
\epsilon_1 {1\over2!} m^2 \; \phi^2(x) + 
\epsilon_2 {\lambda_4 \over {4!}} \; \phi^4 (x) +
\epsilon_3 {\lambda_6 \over {6!}} \; \phi^6 (x) 
\right\}
\nonumber\\
&&\qquad
+O(\hbar \phi (\partial\phi)^2) +O(\hbar^2).
\end{eqnarray} 
%

\section{RGE's and effective action for 
$\lambda(\phi^{{\lowercase{n}}})_2$} 
 
In two dimensions any scalar field theory with polynomial interactions
is renormalizable. The field $\phi$ is canonically dimensionless, and
for the bare action we have
\begin{eqnarray} 
S[\phi]= \int \d^2 x \;  
\left\{
{1 \over 2} \partial_\mu \phi(x) \, \partial^\mu \phi(x) + V(\phi)
\right\}. 
\end{eqnarray} 
Here $V(\phi)$ is an arbitrary polynomial in the field $\phi$. All the
coefficients in this polynomial have the {\em same} canonical
dimension, that of $m^2$. The divergent terms of the effective action
in $2$ dimensions are
\begin{eqnarray} 
\Gamma_1^\Lambda [\phi]&=&  
 -{1 \over {4\pi}} \ln{\Lambda\over m} 
\int \d^2 x \; 
\left\{ V''(\phi(x)) - m^2 \right\}
+ O(\Lambda^{-2}). 
\end{eqnarray} 
Where we have defined $m^2 = V''(\phi=0)$.  Note that there is now
only a logarithmic divergence. The wavefunction renormalization again
vanishes to one loop, and we have
\begin{eqnarray} 
\Gamma[\phi]&=& S[\phi] -S[0] + \hbar\Gamma_1[\phi] +O(\hbar^2),
\end{eqnarray}
with
\begin{eqnarray}
\Gamma_1[\phi]&=& \Gamma_1^\Lambda[\phi]+ \Gamma_1^{\mathrm{finite}}[\phi]. 
\end{eqnarray} 
This is a renormalizable theory, and therefore we can split the bare
parameters into renormalized parameters and counterterms and write
\begin{eqnarray} 
S[\phi]
&=&   
S[\phi, \mu] 
+{\hbar \over {4\pi}} \ln{\Lambda\over\mu} 
\int \d^2 x \; 
\left\{ V''(\phi(x)) - m^2 \right\}
+ O(\hbar \Lambda^{-2}) + O(\hbar^2), 
\end{eqnarray} 
so that the effective action can be rewritten as 
\begin{eqnarray} 
\Gamma[\phi]&=& S[\phi, \mu] - S[0]
-{\hbar \over {4\pi}} \ln{\mu\over m} 
\int \d^2 x \; 
\left\{ V''(\phi(x)) - m^2 \right\}
+\hbar \Gamma_1^{\mathrm{finite}}[\phi] 
+ O(\hbar\Lambda^{-2})
+O(\hbar^2).
\end{eqnarray} 
The corresponding renormalization group equations are most easily
collected together and presented as
\begin{equation}
\mu {\d V(\phi(x),\mu)\over\d\mu} = 
{\hbar \over {4\pi}}
\left\{ V''(\phi(x),\mu) - m^2(\mu) \right\} 
+O(\hbar^2).
\end{equation}
Integrating yields
\begin{equation}
V(\phi(x),\mu) = V(\phi(x),\mu_0) +
{\hbar \over {4\pi}}
\left\{ V''(\phi(x),\mu_0) - m^2(\mu_0) \right\} \ln{\mu\over\mu_0}
+O(\hbar^2).
\end{equation}
We can use this to renormalization group improve the classical
potential, but the reconstruction technique we have applied previously
does not enable us to recover the full one-loop effective
potential. The best we can do with the reconstruction technique is to
obtain
\begin{equation}
{\cal V}(\phi_0,\mu) = V(\phi_0,\mu_0) +
{\hbar \over {4\pi}}
\left\{ V''(\phi_0,\mu) - m^2(\mu) \right\} \ln{\mu\over m(\mu)}
+\hbar m^2(\mu) \; X[V(\phi_0,\mu)/m^2(\mu),\mu]
+O(\hbar^2).
\end{equation}
Here $X[(V(\phi_0,\mu)/m^2(\mu),\mu]$ is an arbitrary dimensionless
function of the indicated variables. Unfortunately, since $\phi_0$ is
now dimensionless we cannot use dimensional analysis to further pin
down the behaviour of $X$. To reduce the remaining freedom, it is best
to backtrack to our earlier result (\ref{E:homogeneous}) which for two
dimensions yields
\begin{equation}
{\cal V}(\phi_0,\mu) = V(\phi_0,\mu_0) +
{\hbar \over {8\pi}}
\left\{ V''(\phi_0,\mu) \ln{V''(\phi_0,\mu)\over\mu^2}
- m^2(\mu) \ln{m^2(\mu) \over\mu^2} \right\} 
+O(\hbar^2),
\end{equation}
up to finite renormalizations. This latter result can now be inserted
into what we know about the effective action to determine (up to
finite renormalizations)
\begin{eqnarray}
\Gamma[\phi(x)] 
&=& 
S[\phi(x),m(\mu),\lambda(\mu)] + 
\int \d^2x \Bigg[
{{\hbar} \over{8\pi} } 
\left\{ 
V''(\phi(x),\mu) 
\ln{ V''(\phi(x),\mu) \over \mu^2}
- m^2(\mu) \ln { m^2(\mu) \over \mu^2} 
\right\}
\nonumber
\\
&&
\qquad \qquad \qquad
+ O(\hbar \phi^2 (\partial\phi)^2)  + O(\hbar(\partial\phi)^4) \Bigg]
+ O(\hbar^2).
\end{eqnarray} 
This finally gives us the leading term in the gradient expansion for
the one-loop effective action in $\lambda (\phi^n)_2$.

\section{Discussion} 

In this paper we have treated the topic of renormalization
group improved perturbation theory using a functional-integral based
calculation of the (1-loop) effective action and/or effective
potential for renormalizable scalar quantum field theories. We have
sought to stress the importance of renormalization-group improved
perturbation theory and the equivalence in leading logarithms between
the improved action and the effective action computed by standard
(path integral or Feynman diagram) means.

The need for such an article addressing the topic of improved
perturbation theory is motivated by the lack of any adequate coverage
of this fundamental concept in the literature. Thus, while aspects of
improved perturbation theory may be known as ``folklore'' to many
practitioners in the field, there is nonetheless no accessible
reference regarding RG improvements of the action (or Lagrangian, or
even the potential).

But apart from attempting to repair this serious omission, the purpose
of this article is really two-fold: firstly, to demonstrate the ease
with which the one-loop renormalization group equations may be
obtained working directly from only the divergent part of the one-loop
effective action, totally side-stepping the need to handle Feynman
diagrams. This is a ``plus'', since it is frequently much easier to
calculate and identify the divergences in the effective action than it
is to use perturbation theory to calculate beta-functions. Secondly,
if we have prior knowledge of the renormalization group equations, the
solutions of these may be immediately used to improve all tree-level
expressions to yield valuable information of the leading-log terms
that {\em must} be present in the same expressions calculated to
one-loop. The point is, we do not have to calculate the leading-log
corrections of these expressions directly: we simply {\em improve}
the tree-level expression by substituting the bare parameters
appearing in them by their running scale-dependent forms.  This holds
of course not only for the action and potential, but also for Green
functions, cross sections, scattering amplitudes, couplings, {\em etc.}.

Improvement consists of substituting the bare parameters appearing in
a (renormalizable) field theory by their running, scale-dependent
forms, calculated to some order in the coupling(s). The very scale
dependence of these couplings is of course a consequence of the
fluctuations and interactions present in the theory. The scale
dependence is handled by the renormalization group whose aim is to
describe how the dynamics of a system evolves as we change the scale
at which the phenomena is being observed. Improved perturbation theory
then results from combining the tools of renormalization group with
perturbation theory and allows us to go beyond the strict limitations
imposed by conventional perturbation theory alone.

When cross-fertilized with zeta function technology, the ideas
presented in this article provide an exceptionally clean and
compact formalism for extracting all of one-loop physics from an
appropriate Seeley-DeWitt coefficient. Details will be presented
elsewhere~\cite{HMPV-a2}.

Applications of these powerful techniques are by no means limited to
``just'' quantum field theory. A vast range of non-linear physical
phenomena subject to fluctuations, not necessarily of a quantum
mechanical nature, may also be investigated using much of the same
technology already developed for quantum field theory. We have in mind
processes subject to thermal or statistical noise which abound in
phenomenology ranging from, but not limited to, the problem of pattern
formation, convection and hydrodynamic turbulence, chaos, chemical
instabilities, and morphogenesis~\cite{Cross-Hohenberg,Frisch}. All
such phenomena can be modelled by non-linear reaction-diffusion
equations of one rubric or another, with non-potential and/or
derivative interactions as well as with conventional polynomial
potentials. The inevitable stochasticity inherent in these systems can
be incorporated by means of a noise source, and we have to consider in
general stochastic nonlinear parabolic equations. The dynamics encoded
in these equations can be equivalently and profitably re-cast in terms
of generating functional integrals, thus converting stochastic
dynamics {\em per se} into a field-theory language that can be
calculationally exploited in a maximally efficient
manner~\cite{HMPV-sde,HMPV-kpz}.

A most interesting formal analogy that arises between quantum and
stochastic field theories is in the identification of a loop-counting
parameter.  It turns out that at least for processes subject to white
Gaussian noise, the modulus of the noise two-point function {\em is}
the loop-counting parameter in a field theory formulation of
statistical fluctuations, and this is the analog of Planck's constant
$\hbar$ which is the loop counting parameter when the fluctuations are
of a quantum nature. But there is much more. As the system is subject
to fluctuations, we can expect the parameters appearing in the
stochastic equations to run with distance or momentum scale as
dictated by corresponding renormalization group equations. These
equations can be just as simply calculated for stochastic field theory
as they are for quantum field theory by way of the effective action in
the manner shown here.  And, just as in the case of quantum field
theory, their solutions can be used to improve expressions based on
tree-level stochastic equations. We shall report on these developments
elsewhere~\cite{HMPV-sde,HMPV-kpz,HMPV-rdd}.

\section*{Acknowledgments} 

This work was supported in Spain by the Spanish Ministry of Education
and Culture and the Spanish Ministry of Defence (DH and JPM). In the
US, support was provided by the US Department of Energy (CMP and
MV). Additionally, MV wishes to acknowledge support from the Spanish
Ministry of Education and Culture through the Sabbatical Programme,
and to recognize the kind hospitality provided by LAEFF (Laboratorio
de Astrof\'\i{}sica Espacial y F\'\i{}sica Fundamental; Madrid,
Spain). Finally, MV also wishes to thank Victoria University (Te Whare
Wananga o te Upoko o te Ika a Maui; Wellington, New Zealand) for
hospitality during final stages of this work.

 

\begin{references}
\bibitem{Textbooks}
The number of textbooks on quantum field theory has grown explosively
over the last few years. Apart from the classics
\cite{Bjorken-Drell-1,Bjorken-Drell-2,Landau-Lifshitz,Bogolubov-Shirkov},
and up to date soon to be classics~\cite{Weinberg-1,Weinberg-2} there
are now textbooks available at almost every level of detail and
clarity, from general surveys to specialized topics. Despite the wide
range of topics and numerous techniques covered in the books
from~\cite{Nash} to~\cite{Zinn-Justin}, the material of this article
has (inexplicably to us) not been treated elsewhere.
\end{references}
\end{document}